\documentclass[aps,twocolumn,superscriptaddress,floatfix]{revtex4}%
\usepackage{graphicx}
\usepackage{bm, amsmath, amssymb}
\usepackage{color}
\usepackage{soul}
\usepackage{amstext}
\usepackage{amsmath}
\usepackage{amsfonts}
\usepackage{amssymb}
\usepackage[T1]{fontenc}%
\usepackage{hyperref}

\usepackage{dcolumn}
\usepackage{cleveref} 
\usepackage{braket}
\usepackage{subfigure}
\usepackage{makecell}
\usepackage{tabularx}
\usepackage{array}

\begin{document}
\title{All-Microwave Multiqubit Gates}
\author{Guanqi Wang}
\affiliation{Ministry of Education Key Laboratory for Nonequilibrium Synthesis and Modulation of Condensed Matter, Shaanxi Province Key Laboratory of Quantum Information and Quantum Optoelectronic Devices, School of Physics, Xi'an Jiaotong University, Xi'an 710049, China}
\affiliation{Shenzhen International Quantum Academy, Futian District, Shenzhen, P. R. China}
\author{Yao Song}
\affiliation{Shenzhen Institute for Quantum Science and Engineering, Southern University of Science and Technology, Shenzhen, P. R. China}
\affiliation{Shenzhen International Quantum Academy, Futian District, Shenzhen, P. R. China}
\author{Peng-Bo Li}
\affiliation{Ministry of Education Key Laboratory for Nonequilibrium Synthesis and Modulation of Condensed Matter, Shaanxi Province Key Laboratory of Quantum Information and Quantum Optoelectronic Devices, School of Physics, Xi'an Jiaotong University, Xi'an 710049, China}
\author{Xiu-Hao Deng}
\thanks{Corresponding author: dengxiuhao@iqasz.cn}
\affiliation{Shenzhen International Quantum Academy, Futian District, Shenzhen, P. R. China}
\affiliation{Shenzhen Branch, Hefei National Laboratory, Shenzhen, 518048, China}

\begin{abstract}
Implementing high-fidelity multiqubit gates is critical for reducing circuit depth in near-term quantum processors and fault-tolerant architectures. With the realization of multiqubit interactions and entanglement remains a critical challenge that limits the implementation of multiqubit gates. Here we propose an all-microwave scheme to realize single-step multiqubit gates—incorporating $n$ control and $m$ target qubits—tailored for frequency-tunable superconducting transmon networks. By leveraging cross-resonance (CR) drives, this approach induces effective two-body $ZX$ interactions that are significantly stronger than those achieved in conventional resonant regimes, circumventing the need for tunable couplers. The system's effective Hamiltonian is analytically derived using both quad frame rotation and rigorous block diagonalization. These combined theoretical methods facilitate the identification of optimal parameter regimes that simultaneously enhance desired target couplings and suppress parasitic higher-order terms, such as $ZXX$ interactions. Through numerical simulations and gradient-based pulse shape optimization, we demonstrate three-qubit gates achieving fidelities exceeding $99.9\%$ within a $60$ ns gate duration. Furthermore, we evaluate the fundamental scalability of this framework by extending the optimization to a five-qubit CXXXX architecture, confirming the physical viability of the multi-target driving scheme. Because these multiqubit operations naturally emulate stabilizer measurements, this architecture provides a hardware-efficient strategy for reducing the circuit depth of stabilizer-type operations by replacing sequential two-qubit gate decompositions with single-step multiqubit gates.

\end{abstract}

\maketitle

\section{Introduction}
In the NISQ era ~\cite{huang2020superconducting,horowitz2019quantum}, practical quantum computing is constrained by limited gate fidelity and restricted circuit scalability ~\cite{bharti2022noisy,kechedzhi2018blueprint}, which significantly hinder the implementation of quantum algorithms. A promising approach to mitigate those limitations is the use of multiqubit gates, which reduces circuit depth compared with decompositions into single- and two-qubit operations. While single- and two-qubit gates have been extensively studied with well-established designs, including numerous realizations across superconducting circuits~\cite{xu2020high}, spin qubits~\cite{xie202399}, and trapped ions ~\cite{hughes2025trapped,ni2025swap}, multiqubit gate research remains relatively underdeveloped. Existing two-qubit implementations exploit diverse interaction mechanisms~\cite{abughanem2024two} and qubit modalities~\cite{bartling2025universal,nagerl2000ion}, such as transmon and qutrit systems ~\cite{zhao2022quantum,baekkegaard2019realization}, and employ techniques including fixed coupling~\cite{goss2024extending}, coupler-mediated interactions~\cite{mitchell2021hardware}, and tailored driving schemes~\cite{wei2022hamiltonian} for Hamiltonian engineering in superconducting platforms ~\cite{di2022extensible,krantz2019quantum}. In addition, a variety of analytical and numerical methods for dynamical simulation and pulse optimization have enabled the realization of arbitrary two-qubit gates in SU(4) ~\cite{chen2025efficient,sugawara20254}.

To address the constraints in the NISQ era and exploit multi-interaction dynamics~\cite{khazali2020fast,liu2025direct}, substantial efforts have been devoted to multiqubit gate design ~\cite{zahedinejad2015high,rasmussen2020single}, leading to several recent advances ~\cite{abughanem2025practical,wang2025time}. For instance, a high-fidelity iToffoli gate has been demonstrated in superconducting systems using cross-resonance (CR) drives to induce strong two-body ZX interactions ~\cite{kim2022high}. Additionally, a high-fidelity CCZ gate has been realized by employing two tunable couplers among three qubits ~\cite{liu2025direct}. Motivated by these developments, we propose a generalized framework for constructing multiqubit gates with n control and m target qubits via CR-driven two-body couplings, enabling single-step multiqubit operations and reducing circuit depth compared with conventional two-qubit decompositions.

The CR effect, first proposed in ~\cite{rigetti2010fully}, has become a key method for engineering multi-body interactions and realizing target unitaries. Analytical approaches for deriving effective Hamiltonians have been extensively developed ~\cite{magesan2020effective,khazali2020fast}, and CR drives are now widely employed to implement control–target (ZX) interactions ~\cite{gorshkov2025cavity,patterson2019calibration,ohfuchi2024remote}. Quantum gate design typically leverages platform-specific features, and in superconducting systems, CR-based techniques are particularly effective.
\begin{figure}[htp]
\centering
\includegraphics[width=8.5cm]{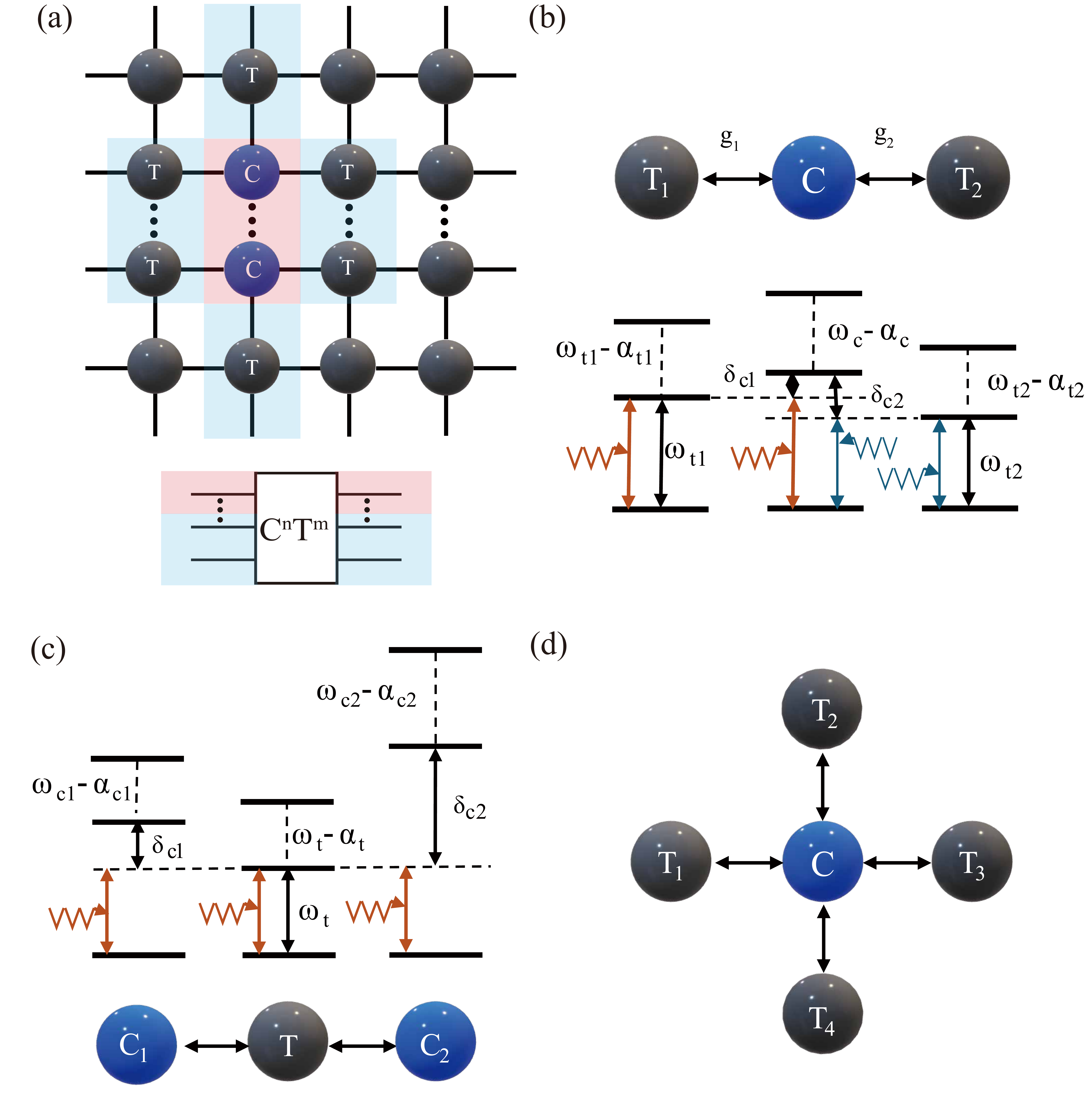}
\caption{Framework of the proposed model. (a) Schematic of the (n+m) multiqubit gate on a superconducting qubit network, where each circle represents a qubit, black circles (T) denote target qubits and blue circles (C) denote control qubits. The bottom part illustrates how the structure serves as a resource in quantum circuits. The pink and blue backgrounds represent the control and target qubits, respectively. (b) This figure illustrates the model for designing a three-qubit gate. The three qubits are coupled with strengths $g_1$ and $g_2$, and their energy levels are shown on the bottom. Two CR drives resonant with $\omega_{t1}$ and $\omega_{t2}$ are applied to the control qubit, while two additional resonant drives are applied to the target qubits to induce local control. (c) Three-qubit CCX (Toffoli) gate, where the central qubit acts as the target and is coupled to two control qubits. (d) Extension to larger systems, illustrated by a CXXXX gate.}
\label{model001}
\end{figure}

In this work, we present a coupler-free, all-microwave framework for implementing a broad family of (n+m) controlled multiqubit gates with Pauli-type target operations in frequency-tunable superconducting transmon networks, providing a proof-of-principle route toward scalable architectures. In Sec.\uppercase\expandafter{\romannumeral2}, we introduce the system models and describe the design of fixed couplings and CR drives. In Sec.\uppercase\expandafter{\romannumeral3}, we derive the effective Hamiltonian using two analytical approaches: quad frame rotation, which provide intuitive physical insight ~\cite{zeuch2020exact,fujii2013introduction}, and block diagonalization, which captures a broader range of interactions beyond the dominant ZX term ~\cite{magesan2020effective}. Numerical simulations are also presented to support the analysis. In Sec.\uppercase\expandafter{\romannumeral4}, high-fidelity gate performance is demonstrated through pulse optimization. Finally, in Sec.\uppercase\expandafter{\romannumeral5}, we assess the experimental feasibility of the proposed scheme by examining distinct target frequencies, control-pulse bandwidths, and sensitivity to variations in transmon parameters. The resulting multiqubit gates exhibit stabilizer-like properties and can substantially reduce circuit depth compared with implementations based on conventional single- and two-qubit gate decompositions~\cite{2025Scalable}, making them suitable for applications such as quantum error correction and quantum algorithm optimization. Moreover, the coupler-free design reduces fabrication complexity. These advancements provide a promising pathway toward scalable quantum processor architectures and efficient quantum circuit implementations. 

\section{Construction of Multiqubit CR Gates}
This section introduces a generalized framework for designing multiqubit (n+m) gates, where the target operations include (X), (Y), and (Z), using all-microwave CR drives in frequency-tunable transmon systems with XX couplings. We first describe our regime through two basic models with representative examples and subsequently extend it to systems with larger numbers of qubits. The design rationale, supported by detailed analytical calculations, is presented in the following section. Our approach enables the realization of a family of (n+m) gates, including, for example, the Toffoli gate ~\cite{he2017decompositions,dong2024experimental}. The CR regime involves applying a drive that is off-resonant with the qubit’s transition frequency but resonant with a coupled qubit, distinguishing it from conventional resonant driving. Initially proposed in ~\cite{rigetti2010fully}, the CR effect has been experimentally validated as a practical method for constructing multiqubit gates ~\cite{kim2022high}. As illustrated in Fig.~\ref{model001}(a), the framework of the (n+m) gate model consists of qubits represented as nodes, where those labeled T denote target qubits and those labeled C denote control qubits. The bottom part presents the n control m target configuration as an example, demonstrating how such a structure can serve as a resource in quantum circuits. The pink and blue backgrounds represent the control and target qubits, respectively.\par Now we are going to introduce two basic units in building our system-three qubit CTT and CCT. Fig.~\ref{model001}(b) shows the explicit model for CTT gate (with (T $\in$ {X,Y,Z})). As a specific example, we consider the CXX (taking T as X for simplicity here) case. Two CR drives, resonant with the target qubits, are applied to the control qubit to induce many-body interactions, while additional resonant drives are applied to the target qubits to generate local Hamiltonian terms. These CR drives induce two independent ZX interactions between the control and each target qubit. Notably, this design does not impose strict constraints on coupling strengths or qubit frequencies, enhancing its experimental feasibility. We now briefly outline the model. The Hamiltonian is given by
\begin{equation}\begin{aligned}
H=&\frac{\omega_{t1}}{2}\sigma_{t1}^z+\frac{\omega_{t2}}{2}\sigma_{t2}^z+\frac{\omega_c}{2}\sigma_c^z+\frac{g_1}{2}\sigma_{t1}^x\sigma_{c}^x+\frac{g_2}{2}
\sigma_c^x\sigma_{t2}^x\\
&+\Omega_{t1}\text{cos}(\omega_{t1}t+\phi_{t1})\sigma_{t1}^x+\Omega_{t2}\text{cos}(\omega_{t2}t+\phi_{t2}^x)\sigma_{t2}^x\\&+[\Omega_{c1}\text{cos}(\omega_{t1}t+\phi_c)+\Omega_{c2}\text{cos}(\omega_{t2}t+\phi_c)]\sigma_c^x\label{1}\end{aligned}\end{equation}
Hereafter, we denote the Pauli operators \(\sigma_x\), \(\sigma_y\), and \(\sigma_z\) equivalently by \(X\), \(Y\), and \(Z\), respectively.
Here, the dynamics of the two control–target pairs remain effectively decoupled, and for simplicity, only two levels are taken into account. A model incorporating more levels will be considered in a later section to complete the description. In the dispersive regime ($\Delta_i=|\omega_c-\omega_{ti}|\gg g_i$), the effective Hamiltonian reduces to
\begin{equation}\begin{aligned}
H^{eff}&=\frac{\Omega_{t1}}{2}\text{cos}\phi_{t1}\sigma_{t1}^x+\frac{\Omega_{t1}}{2}\text{sin}\phi_{t1}\sigma_{t1}^y+\frac{\Omega_{t2}}{2}\text{cos}\phi_{t2}\sigma_{t2}^x\\&+\frac{\Omega_{t2}}{2}\text{sin}\phi_{t2}\sigma_{t2}^y+\frac{g_1\Omega_{c1}}{4\delta_{c1}}\sigma_c^z\sigma_{t1}^x+\frac{g_2\Omega_{c2}}{4\delta_{c2}}\sigma_c^z\sigma^x_{t2}\label{2}.
\end{aligned}\end{equation}
(See Appendix~\ref{Appen_Heff1}, Eqs.~(\ref{r1})-(\ref{r2}) for more details.)
By applying a composite pulse to the control qubit simultaneously, formed by the coherent superposition of the two CR drives, both control–target entangling operations can be executed in parallel. This approach has been experimentally demonstrated on superconducting quantum processors ~\cite{2020Quantum}. However, although experimentally feasible, this model is analytically complex. Therefore, to simplify the theoretical analysis we take some specific parameters into account. The frequencies of the target qubit are taken to be the same and only one CR pulse at frequency $\omega_t$ is applied to the control qubit, resonant with the target qubits, thereby inducing effective ZX interactions between the control and each target. In addition, resonant pulses applied to the target qubits implement local (X) operations. The Hamiltonian then becomes \begin{equation}\begin{aligned}
H=&\frac{\omega_t}{2}\sigma_{t1}^z+\frac{\omega_t}{2}\sigma_{t2}^z+\frac{\omega_c}{2}\sigma_c^z+\frac{g_1}{2}\sigma_{t1}^x\sigma_{c}^x+\frac{g_2}{2}
\sigma_c^x\sigma_{t2}^x\\
&+\Omega_{t1}\text{cos}(\omega_tt+\phi_{t1})\sigma_{t1}^x+\Omega_{t2}\text{cos}(\omega_tt+\phi_{t2}^x)\sigma_{t2}^x\\&+\Omega_c\text{cos}(\omega_tt+\phi_c)\sigma_c^x\label{111}\end{aligned}\end{equation}


Fig.~\ref{model001}(c) depicts a configuration with two control qubits (CCX) which serves as another basic model, and the central qubit acts as the target. The control qubits are not required to have identical frequencies, and multiple microwave drives resonant with the target frequency are applied to realize the desired interactions.

Now we consider systems with more qubits, they can be taken as extensions of these two basic models by adding more control or target qubits to the two basic models. As shown in Fig.~\ref{model001}(d), the CXXXX configuration serves as an example in which a central control qubit is coupled to four target qubits. Drives applied to each qubit operate at frequency $\omega_t$, satisfying the CR condition. For clarity, we summarize the core conditions for designing CR-based $(n+m)$ gates within our regime with specific parameters:
\begin{enumerate}
    \item No direct coupling is introduced between qubits of the same type (control–control or target–target).
    \item Couplings are established exclusively between control and target qubits to enable conditional evolution.
    \item All target qubits share a common frequency $\omega_t$, and control drives are applied at this frequency to realize CR-induced interactions and local term engineering; control qubit frequencies may be chosen arbitrarily.
    \item Due to hardware constraints in superconducting systems, each qubit is limited to a finite number of nearest-neighbor connections.
\end{enumerate}

Those specific parameters are introduced to facilitate the theoretical construction of (n+m) gates. The associated constraints can be relaxed in experimental implementations, ensuring that the proposed framework remains feasible and compatible with realistic physical systems.

\section{Analytical Calculation of Multiqubit Gates}
In this section, we present a detailed derivation of the effective Hamiltonian using two complementary approaches. The first is based on rotating-frame transformations, which provide clear physical insight into the underlying dynamics. The second employs block diagonalization, offering higher accuracy and enabling the treatment of two- and three-body interactions beyond the two-level approximation for transmon qubits. This approach facilitates the identification of optimal parameters that simultaneously suppress undesired interactions and enhance target interactions, thereby supporting efficient numerical optimization.

\subsection{Effective Hamiltonian by Quad Frame Rotation}

We first derive the effective Hamiltonian via quad frame rotation and demonstrate the design of (n+m) gates based on the resulting interactions, using CXX and CCXXX gates as representative examples as they represent the two basic models and the extension with more qubits. The derivations for other multiqubit gates follow analogous procedures (see Appendix~\ref{Appen_Heff1}, Eqs.~(\ref{r3})-(\ref{r4}) for further details).

As an illustrative case, we consider the CXX gate, the derivation for other gates proceeds similarly. The model, shown in Fig.~\ref{model001}(b), consists of three qubits, where the central qubit is coupled to the two targets with coupling strengths $g_1$ and $g_2$ (assumed equal for simplicity). The control qubit is driven by $\Omega_c(t)\text{cos}(\omega_tt+\phi_c)$, while the target qubits are driven by $\Omega_{t1}(t)\text{cos}(\omega_tt+\phi_{t1})$ and $\Omega_{t2}(t)\text{cos}(\omega_tt+\phi_{t2})$, respectively. The system Hamiltonian is given by Eq.~(\ref{111}). This Hamiltonian is transformed into an effective form through a sequence of rotating-frame transformations, $H'=U^{-1}HU-iU^{-1}\frac{dU}{dt}$ with U represents the rotating frame operator, while rapidly oscillating terms are neglected under the rotating wave approximation (RWA) ~\cite{ansari2019superconducting,scully1997quantum}. The resulting time-independent Hamiltonian exhibits the desired strong ZX interactions, replacing the original XX coupling:
\begin{equation}\begin{aligned}
H^{eff}&=\frac{\Omega_{t1}}{2}\text{cos}\phi_{t1}\sigma_{t1}^x+\frac{\Omega_{t1}}{2}\text{sin}\phi_{t1}\sigma_{t1}^y+\frac{\Omega_{t2}}{2}\text{cos}\phi_{t2}\sigma_{t2}^x\\&+\frac{\Omega_{t2}}{2}\text{sin}\phi_{t2}\sigma_{t2}^y+\frac{g_1\Omega_c}{4\delta_c}\sigma_c^z\sigma_{t1}^x+\frac{g_2\Omega_c}{4\delta_c}\sigma_c^z\sigma^x_{t2}\label{222}
\end{aligned}\end{equation}

While this effective Hamiltonian shows that the local X and Y terms are controlled by the drive phases, the independent tunability of both local terms is limited by a single phase parameter $\phi_{ti}$. To overcome this constraint, we introduce an extended scheme incorporating additional control degrees of freedom via two extra drives, yielding four pulses on the target qubits: $\Omega_{t1}\text{cos}(\omega_tt+\phi_{t1})$, $\Omega_{t2}\text{cos}(\omega_tt+\phi_{t2})$, $\Omega_{t1}'\text{cos}(\omega_tt+\phi'_{t1})$, and $\Omega_{t2}'\text{cos}(\omega_tt+\phi_{t2}^{'})$. The effective Hamiltonian then becomes 
\begin{equation}\begin{aligned}
H^{eff}&=\sum_{{t_1,t_2}}\frac{\Omega_{ti}}{2}[\text{cos}(\phi_{t1})\sigma_{ti}^x+\text{sin}(\phi_{ti})\sigma_{ti}^y]\\&+\sum_{{t_1,t_2}}\frac{\Omega^{'}_{ti}}{2}[\text{cos}(\phi^{'}_{t1})\sigma_{ti}^x+\text{sin}(\phi^{'}_{ti})\sigma_{ti}^y]\\&+\frac{g_2\Omega_c}{4\delta_c}\sigma_c^z\sigma_{t2}^x+\frac{g_1\Omega_c}{4\delta_c}\sigma_c^z\sigma^x_{t1}
\end{aligned}\label{333}\end{equation}
These drives generate additional local X and Y contributions. By choosing $\phi_{t2}=\phi_{t1}=0$ and $\phi_{t2}'=\phi_{t1}'=\frac{\pi}{2}$, the X and Y terms in the effective Hamiltonian become independently tunable. (This can also be taken as add another $\sigma_y$ control on the target qubit) Consequently, all parameters can be controlled individually through the amplitudes of the microwave drives.
Therefore, the original Hamiltonian is mapped to a time-independent effective Hamiltonian characterized by ZX interactions, which are naturally suited for constructing controlled quantum gates such as CXX gate. The detailed construction of the target unitary is provided in Appendix ~\ref{Appen_Heff}, Eqs.~(\ref{r5})-(\ref{r6}). This framework enables the realization of three-qubit gates with independently selectable Pauli-type target operations (X, Y, Z), up to local phase factors that can be readily compensated.\par

Owing to the ability of each transmon qubit to couple with up to four neighboring qubits, the scheme extends naturally to larger systems. For example, the CXXXX gate shown in Fig.~\ref{model001}(d) operates under the same principles as the CXX gate, with additional target qubits coupled to the central control.

We next consider the construction of the CCXXX gate. The approach follows that used for the Toffoli and CXX gates ~\cite{kim2022high}. Based on the design principles introduced earlier, the corresponding Hamiltonian is
\begin{equation}\begin{aligned}
H&=\sum_{i=1,2}\frac{\omega_{ci}}{2}\sigma_{ci}^z+\sum_{i=1}^3\frac{\omega_t}{2}\sigma_{ti}^z\\&+\sum_{i=c1,c2,t1,t2,t3}[\Omega_i\text{cos}(\omega_tt+\phi_i)+\Omega_i'\text{cos}(\omega_tt+\phi_t')]\sigma_i^x\\&
+\sum_{i=c1,c2.k=1}^{k=6}\sum_{j=t1,t2,t3}\frac{g_k}{2}\sigma_i^x\sigma_j^x
\end{aligned}\end{equation}
This Hamiltonian is transformed into a rotating frame through a four-stage unitary transformation under the RWA,
yielding the effective dynamics
\begin{equation}\begin{aligned}
H&=\alpha X_{t1}+\alpha'Y_{t1}+\beta X_{t2} +\beta' Y_{t2}+\gamma X_{t3}+\gamma' Y_{t3}\\&+\delta_1 Z_{c1}X_{t1}+\delta_2 Z_{c1}X_{t2} +
\delta_3 Z_{c1}X_{t3}+\delta_4 Z_{c2}X_{t1}\\&+ \delta_5 Z_{c2}X_{t2} +\delta_6 Z_{c2}X_{t3}\label{ac}
\end{aligned}\end{equation}
The three target qubits form a linear superposition, allowing their dynamics to be treated independently. This enables the realization of an X operation conditioned on the control state $\ket{11}$, and the CCXXX gate is achieved by selecting an appropriate gate time $\tau$.

Furthermore, the effective-Hamiltonian framework suggests a possible extension to controlled rotations with arbitrary rotation angles. To illustrate this theoretical possibility, consider the effective interaction \(H_{\mathrm{eff}}=J_{ZX}Z_cX_t\), which generates the unitary evolution
\begin{equation}
\begin{aligned}
U(T) &= e^{-iH_{\mathrm{eff}}T} \\
&= |0\rangle\langle0|_c\otimes e^{-iJ_{ZX}TX_t}
 + |1\rangle\langle1|_c\otimes e^{+iJ_{ZX}TX_t}.
\end{aligned}
\end{equation}
Using the conventional definition \(R_X(\theta)=e^{-i\theta X/2}\), this evolution corresponds to opposite \(X\)-axis rotations conditioned on the control state, with angles determined by the accumulated interaction \(J_{ZX}T\). Within this effective description, a standard controlled-\(R_X(\theta)\) operation can, in principle, be constructed by combining the conditional evolution with an appropriate unconditional rotation on the target. Adjusting the interaction strength or duration therefore provides a route to continuously tunable rotation angles. An analogous construction applies to controlled-\(R_Y(\theta)\) rotations when an effective \(ZY\) interaction is generated through suitable microwave phase control.

These observations support the theoretical feasibility of programmable-angle controlled rotations within the effective model. However, a complete implementation of such gates in the full multilevel system, including the treatment of residual interactions and leakage, has not been established in the present work. Future work will focus on developing suitable pulse sequences, optimizing their parameters, and quantitatively evaluating the achievable gate fidelities.

Finally, we generalize the framework to larger multiqubit systems. The combination of control–target coupling and CR driving generates ZX interaction terms that mediate conditional operations on target qubits, while resonant drives applied to targets introduce local X and Y terms. Consequently, the gate design reduces to selecting appropriate qubit connectivity, where different connection topologies enable the implementation of a broad class of multiqubit gates.

\subsection{Effective Hamiltonian by Block Diagonalization with Bloch-Brandow Perturbation}
Although the analytical method proposed above effectively demonstrates the desired effective interactions through succinct calculations, it overlooks certain significant interactions due to the approximations involved. The simplified QFR Hamiltonian is derived under the following scale-separation conditions:
\begin{equation}
\begin{aligned}
\frac{\max\{\Omega_{\max},g_{\max}\}}{2\omega_t} &\ll 1,
\qquad
\frac{g_{\max}}{\eta_c} \ll 1,
\qquad
\frac{|\Omega_c|}{|\delta_c|} \ll 1, 
\end{aligned}
\label{eq:qfr_validity}
\end{equation}
Here, $\Omega_{\max}$ and $g_{\max}$ denote the largest relevant drive amplitude and control--target coupling strength, respectively. The first condition justifies neglecting the counter-rotating terms under the rotating-wave approximation (RWA). The second condition supports neglecting the coupling terms oscillating at $\eta_c$ in the subsequent rotated frame, provided that no additional near-resonance invalidates this averaging. The third condition justifies the weak-drive simplification used to neglect the additional transverse interaction terms.

Together, the conditions in Eq.~\eqref{eq:qfr_validity} specify the approximation regime of the simplified QFR Hamiltonian. Within this regime, QFR provides an intuitive and concise description of the dominant interactions. When these conditions are not satisfied, the simplified Hamiltonian is no longer justified by the approximations used in its derivation. To overcome this limitation, we now present a more accurate approach for computing effective interactions utilizing block diagonalization~\cite{2015Effective} , which yields the analytical form of any two- or three-body interactions. Additionally, this methodology accommodates models incorporating higher energy levels, which is particularly essential for realistic simulations of transmon qubits and we take three levels into account in the following derivations (Appendix~\ref{Appen_Bloch}, Eqs.~(\ref{rr1})-(\ref{rr5}).
\begin{figure}[t]
\centering
\includegraphics[width=8.8cm]{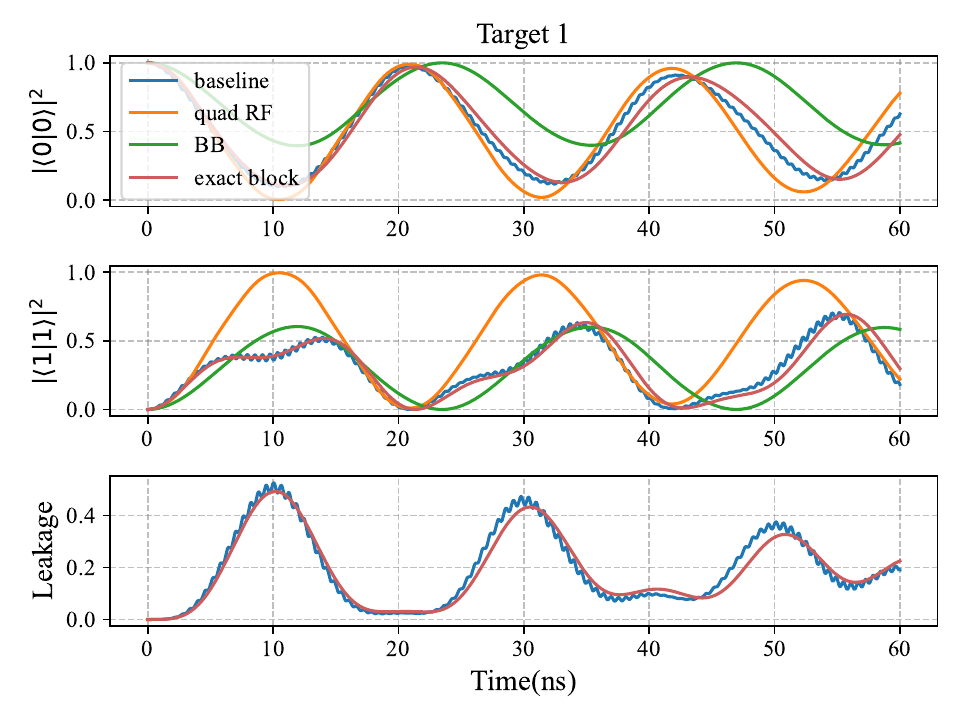}
\caption{The fidelity of different methods is assessed in this figure through dynamic simulations of the system depicted in CXX. Three energy levels are taken into account. Subplots show time evolution of the first, second and leakage level of the target qubit 1. The blue curve represents actual dynamics from the initial Hamiltonian. The orange and the green curves indicate effective Hamiltonian results calculated by quad frame rotation and block diagonalization with Bloch theory. And the red one represents exact block diagonalization.}
    \label{verify}
\end{figure}
This block diagonalization approach offers a highly precise method for deriving the effective Hamiltonian~\cite{cederbaum1989block}, theoretically grounded in the least action principle~\cite{guan2026optimal,li2022nonperturbative}. We first consider exact block diagonalization based on least-action principle for numerical precision and subsequently employ Bloch-Brandow Perturbation theory (BB formalism)~\cite{guan2026optimal} to extract tractable analytical expressions. The effective Hamiltonian is obtained through the following procedure. Since the initial Hamiltonian is time dependent, a RWA is applied, resulting in a time-independent Hamiltonian. Next, this Hamiltonian is transformed into a block diagonal form. Here, $S$ denotes the eigenvector matrix of the Hamiltonian, and $S_{BD}$ represents its block-diagonalized form. The block-diagonalized effective Hamiltonian can then be calculated using this framework.
\begin{equation}
H_{BD}=T^{\dagger}HT , \quad T=SS_{BD}^{\dagger}(S_{BD}S^{\dagger}_{BD})^{-1/2}
\end{equation}
This outcome is entirely determined by the eigenvector matrix $S$ of the RWA Hamiltonian. Consequently, the effective block-diagonalized Hamiltonian can be obtained solely by computing the eigenvector matrix of $H_{RWA}$. Subsequently, the strength of any effective interaction can be evaluated by projecting the matrix onto the corresponding interaction matrix through the trace operation, such as $\operatorname{Tr}[H_{\rm eff}\cdot ZIX]$.

Although the exact block diagonalization method is quite useful numerically, it is impractical for analytical results due to its computational complexity. We therefore employ block diagonalization with Bloch Perturbative theory (BB formalism)~\cite{2015Effective} by following these steps: first, the RWA is used to eliminate the time-dependent terms, yielding the effective Hamiltonian in the computational subspace as $H_{\rm eff}=PH_0P+V_{\rm eff}$, where $P$ is the projection operator. The term $V_{\rm eff}$ is obtained via perturbative expansion:
\begin{equation}
\begin{aligned}
   & V_{\rm eff}^{(1)}=PVP \\
   & V_{\rm eff}^{(2)}=P[V(V)]P \\
   & \vdots
\end{aligned}
\end{equation}
with the superoperator $(V)$ defined by
\begin{equation}
[(V)]_{Ii}=\frac{1}{\xi_i-\xi_I}\bra{I}V\ket{i}.
\end{equation}
Here $\xi_i$, $\xi_I$, $\ket{i}$, and $\ket{I}$ denote the eigenvalues and eigenstates, respectively.

This method is instrumental in identifying optimal parameter regimes that simultaneously suppress undesirable interactions while enhancing the desired ZX couplings. For instance, the three-body ZXZ interaction computed via this method yields a value of zero, while the strength of the XXZ interaction is shown in Fig.~\ref{verifyy}(c). All interactions can be calculated through this method, and several analytical forms are given in Appendix ~\ref{Appen_Bloch}, Eqs.~(\ref{rr5})-(\ref{rr2}). For example, the important two-body ZX and crosstalk ZZ interactions are calculated in the three-level model by this method and are provided in the Appendix.

\subsection{Numerical Simulation and Error Budget}
Fig.~\ref{verify} illustrates the simulation of the first qubit of the CXX model with effective Hamiltonians calculated by the two methods to evaluate their fidelity. The dynamics of the first target qubit are depicted. Since the anharmonicity of superconducting transmons is non-negligible, three energy levels are incorporated. The three subplots in Fig.~\ref{verify} display the population dynamics of the first, second, and leakage levels of the first target qubit, respectively. The actual dynamics are represented by the blue curve, while the orange and green curves represent the outcomes of the two effective-Hamiltonian methods, and the red one represents exact block diagonalization. Thereby demonstrating the effectiveness of our approach for calculating effective interactions and its suitability for designing multiqubit gates and selecting appropriate parameters. Furthermore, optimized pulse shaping can be employed to suppress leakage errors, and well-chosen parameters significantly facilitate the effectiveness of such shape optimization.

\begin{figure}[t]
\centering
\includegraphics[width=8.5cm]{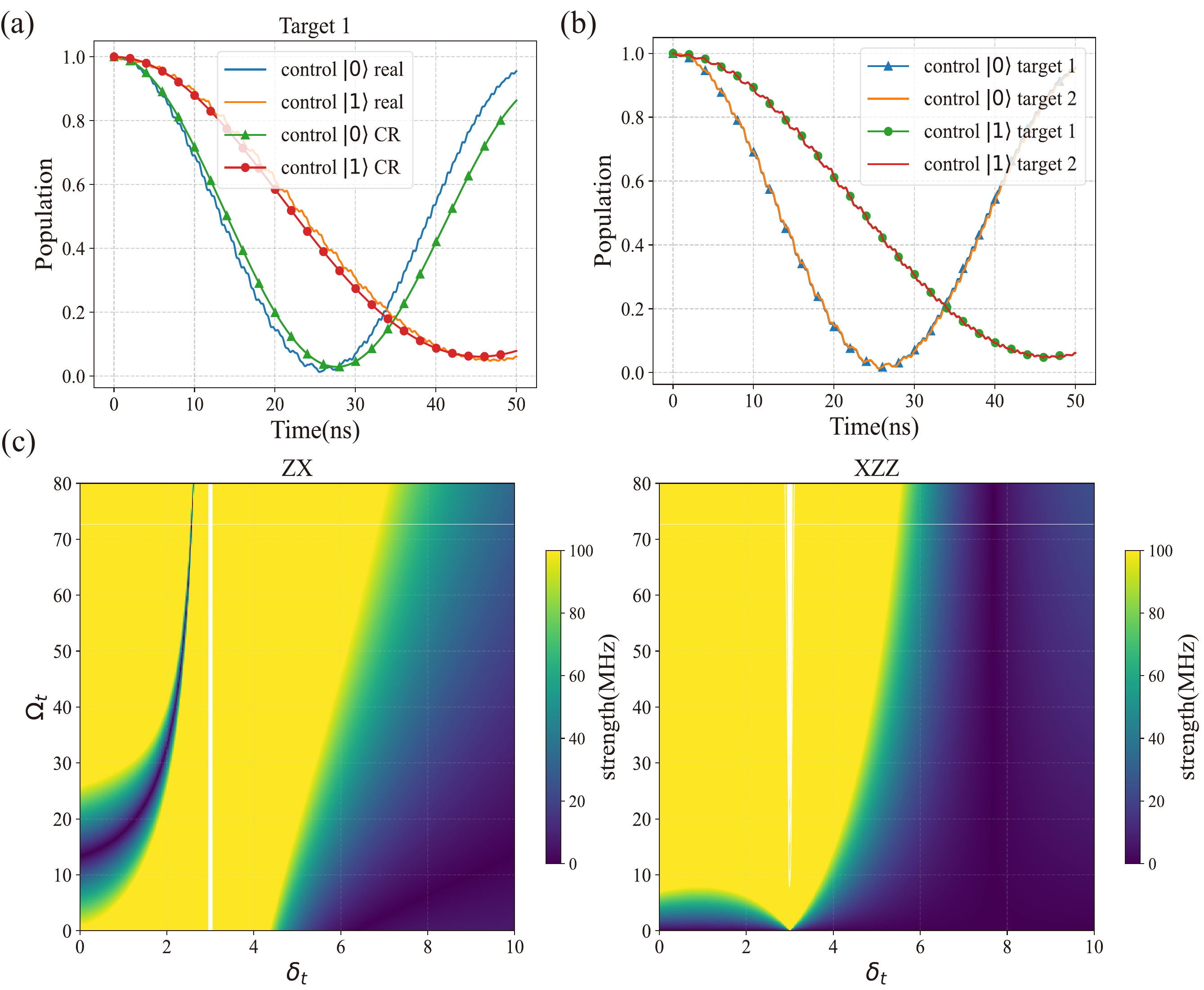}
\caption{(a) Dynamics of the CXX gate under optimized parameters in 50 ns, the blue and orange curves denote the occupation of the first and second target qubit, respectively, when the control is in state $\ket{0}$, while the green and red curves represent the occupations of target 1 and target 2 under control state $\ket{1}$. The dynamics of the two target qubits exhibit similarity, as their evolution can be treated independently. (b) Dynamics of target 1 in CXX: the blue and orange curves correspond to the actual evolution, and the remaining two curves depict the dynamics via effective Hamiltonian. (c) Comparison between the desired ZX interaction and the unwanted XZZ interaction as functions of $\Omega_{c}$ and $\delta_{c}$ in CXX model calculated by block diagonalization.}
    \label{verifyy}
\end{figure}

Fig.~\ref{verifyy}(a) illustrates the time evolution of the state occupations for both target qubits in the CXX gate over 50 ns. The dynamics of the two target qubits exhibit similar profiles, as their evolutions are decoupled and computed independently, a behavior explained in prior sections. Both target qubits are initialized in state $\ket{0}$, and their oscillation periods depend on the control state ($\ket{0}$ or $\ket{1}$), thereby enabling a controlled-X operation and realizing a CXX gate. The fidelity of this gate is 97\% with well-chosen parameters, which demonstrates the feasibility of our regime but remains insufficient for practical quantum circuit applications.

To illustrate that higher fidelity requires optimization of the drive shape, an error budget is introduced. The errors can be roughly divided into three types: leakage error, incoherent error, and coherent error (unwanted interaction) ~\cite{2026Echo}. The incoherent contribution is not considered here. In addition, when the environment is taken into account, the infidelity caused by decay and decoherence can be expressed as
\begin{equation}
F_{\rm depolarized}=\frac{1+p}{2}
\end{equation}
with $p=e^{-\gamma t}$, where $\gamma$ represents the dissipation rate. The fidelity of a quantum gate is calculated by
\begin{equation}
F=\frac{1}{n(n+1)}[\operatorname{Tr}(MM^{\dagger}) + |\operatorname{Tr}(M)|^2] \label{eq:example1}
\end{equation}
with $M=U_0^{\dagger}R$, which is fully determined by the target unitary $U_0$ and the real unitary $R$ \cite{pedersen2007fidelity}. A simple analysis takes the first part of Eq.~(\ref{eq:example1}) as the infidelity caused by leakage error and the remainder as coherent error. Thus, the infidelity caused by leakage in our simulation is
\begin{equation}
\operatorname{In}F_{\rm leakage}=\frac{1}{n(n+1)}[n-\operatorname{Tr}(MM^{\dagger})]\label{eq:example2}.
\end{equation}
Here $n$ is the ideal dimension of the subspace, and the difference between $n$ and $\operatorname{Tr}(MM^{\dagger})$ quantifies the deviation between the ideal subspace and the real subspace after projection. The coherent error is then $\operatorname{In}F_{\rm coherent}=(1-F)-\operatorname{In}F_{\rm leakage}$. The total infidelity in our simulation in Fig.~\ref{verifyy}(a) is 3\%. The infidelity caused by leakage error, calculated using Eq.~(\ref{eq:example2}), accounts for 74\% of this total infidelity, with the remainder being coherent error.

Another method presented in ~\cite{2019Operation} provides an analytical form of the error budget and reveals the relationship between the infidelity and the real evolution matrix $R$. We therefore apply this method to calculate the proportion of each error type again (details in Appendix~\ref{Appen_error}, Eqs.~(\ref{rr3})-(\ref{rr4}). This method constructs a closest matrix $U$ to the real evolution matrix in order to isolate errors that can be readily compensated by adding several single-qubit gates. The fidelity without these simple errors is then given by $F(R,U)$. For both the three- and five-qubit gates, the simple-error contribution is suppressed to a negligible level by the local single-qubit corrections incorporated in our gate design. We therefore do not consider this contribution further in the following error-budget analysis.
 Subsequently, the remaining errors in $F(R,U)$ are divided into leakage error and coherent error arising from imperfect target rotation:
\begin{equation}
1-F(U,R)=(1-F(R,R'))+(1-F(R',U))+\epsilon_{other},
\end{equation}
where $\epsilon_{other}$ denotes other errors, and $R'$ is a matrix close to $U$ that permits arbitrary leakage from $\ket{1}$ to $\ket{2}$ while allowing arbitrary rotation of the target qubits to quantify the leakage contribution. The residual term is defined as
\[
\epsilon_{\mathrm{other}}
=
[1-F(U,R)]
-
[1-F(R,R')]
-
[1-F(R',U)].
\]
Here $\epsilon_{other}$ may include mixed contributions from the interplay between leakage and coherent dynamics, interference between error amplitudes, and higher-order corrections to the additive approximation. Given the real matrix $R$ and the ideal matrix $U_0$, the proportion of each error can be determined. The results are consistent with those obtained from the simpler method above: leakage and coherent errors account for approximately 74\% and 26\% of the total infidelity, respectively, while the residual contribution \(\epsilon_{\mathrm{other}}\) is sufficiently small to be neglected in the three-qubit regime., and the main calculation outcomes are presented in the Appendix. To suppress these errors and achieve higher fidelity, the pulse is further optimized using the pre-optimized parameters in the next section.

Fig.~\ref{verifyy}(b) compares the dynamics of the first target qubit under the real Hamiltonian and the effective Hamiltonian. The effective curves closely match the actual dynamics, demonstrating the effectiveness of our approach. In Fig.~\ref{verifyy}(c), we present the effective interactions of the three-qubit CXX gate as functions of $\Omega_c$ and $\delta_c$, computed via BB formalism. The detailed forms of those interactions are given in the Appendix ~\ref{Appen_Bloch}, Eqs.~(\ref{rr5})-(\ref{rr2}). Taking the ZX and XZZ interactions as examples, this method enables the selection of improved parameters that simultaneously suppress undesirable terms and amplify target interactions. In our approach, we focus on suppressing harmful three-body interactions such as XZZ and ZZX while leveraging the two-body ZX interaction as the fundamental resource for constructing multiqubit gates. It is noteworthy that three-body interactions may also serve as valuable resources for achieving ideal unitaries, offering a distinct strategy for designing multiqubit gates ~\cite{xu2025parity}.

\section{Numerical Optimization}
In this section, we present the fidelity achieved via a constrained pulse optimization algorithm based on auto-differentiation, COCOA~\cite{song2022optimizing}. First, we demonstrate the numerical results for the construction of the CYY gate, followed by those for the CXX and CXZ gates. These results demonstrate that a broad family of controlled Pauli-type multiqubit gates with n control and m target qubits can be realized within our framework. Subsequently, we show the optimization outcomes for five-qubit gates, which demonstrate the scalability of our approach.   
\begin{figure}[htp]
\centering
\includegraphics[width=8.7cm ]{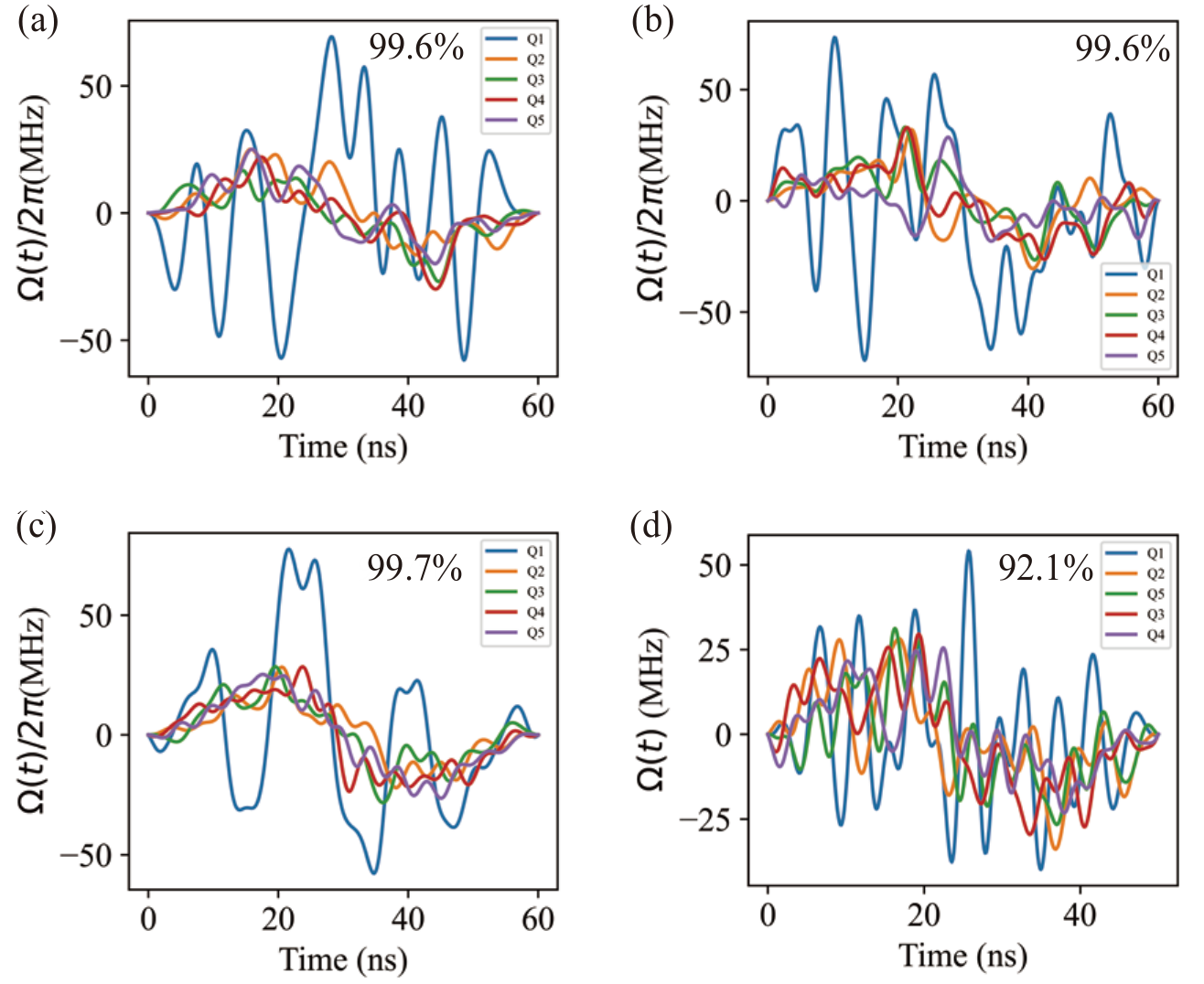}
\caption{Optimization and Fidelity of Multiqubit Gates (a) Optimization of the CYY gate. $\omega_{t1}=5.21\text{ GHz, }\omega_{t2}=5.2\text{ GHz}$, the control qubit frequency is $\omega_c$=5.5\text{ GHz, } and the coupling strengths are $g_{1,2}=8\text{ MHz}$. The anharmonicity is $\alpha_{c}=-300\text{ MHz, }\alpha_{t1}=-320\text{ MHz, }\alpha_{t2}=-350\text{ MHz}$. Five microwave pulses at frequency $\omega_t=5.2\text{ GHz}$ are applied It displays the optimized pulse shapes with Q1 on control and Q2,Q3 on target 1, Q4,Q5 on target 2. And the fidelity after optimization is shown on the top. (b) results for real CXX gates with the same parameters. (c) results for CZX gates.
(d) Optimization of the CXXXX gate. Parameters for this gate are $\omega_{ti(i=1,2...)}=5.2\text{ GHz}$ and $\alpha_i=-300\text{ MHz}$,$\omega_c=5.3\text{ GHz}$ and $g_{1,2...}=8\text{ MHz}$. Five pulses at frequency $\omega_t$ are applied on each qubit. Besides, we also optimize the model for CCXXX gate with a fidelity of 92.6\%.}
\label{opt}
\end{figure}

First, the fidelity of a CYY gate is presented in Fig.~\ref{opt}(a). The CYY gate, which we designed previously,
\begin{equation}
\begin{bmatrix}
1 & 0 & 0&0&0&0&0&0\\
0 & 1 & 0&0&0&0&0&0\\
0 & 0 & 1&0&0&0&0&0\\
0 & 0 & 0&1&0&0&0&0\\
0 & 0 & 0&0&0&0&0&-1\\
0 & 0 & 0&0&0&0&1&0\\
0 & 0 & 0&0&0&1&0&0\\
0 & 0 & 0&0&-1&0&0&0\\
\end{bmatrix}
\label{ai}
\end{equation}
Here, the matrix is expressed in the computational basis \(\{|000\rangle,|001\rangle,|010\rangle,|011\rangle,|100\rangle,|101\rangle,|110\rangle,|111\rangle\}\), where the first qubit is the control qubit and the second and third qubits are the target qubits. Accordingly, the CYY matrix has the block-diagonal form \(CYY=I_4\oplus(Y\otimes Y)\), which gives the nonzero-element pattern shown in Eq.~(\ref{ai}).\par
Since \(Y=SXS^\dagger\), with \(S=\mathrm{diag}(1,i)\), the CYY gate is locally equivalent to the standard CXX gate through single-qubit phase gates on the two target qubits, explicitly given by \(CYY=(I\otimes S\otimes S)CXX(I\otimes S^\dagger\otimes S^\dagger)\). After executing our optimization algorithm, a pulse shape yielding higher fidelity is obtained, and the results are illustrated in the figure. The two pulses applied to the same target offer additional tunability over parameters, so five pulses are shown in the figure. The results demonstrate that the fidelity of our CYY gate can numerically exceed 99.6\% after 200 iterations, which is labeled at the top of the figure, affirming the feasibility of our approach. Moreover, our optimization algorithm can concurrently refine other parameters of the model. The result of the CXX gate is shown in Fig.~\ref{opt}(b). This outcome indicates that the local phase can be effectively compensated through pulse optimization. We further present the results of CZX in Fig.~\ref{opt}(c) to illustrate the universality of our method in designing multiqubit gates. As the model structure can be applied to implement both Z and X(Y) operations, as established in earlier derivations. The results confirm that CZX gates can be realized with fidelities exceeding 99.7\% after 300 iterations. Additionally, the successful implementation of the Toffoli gate (CCX) and CCZ gate aligns well with previous studies \cite{kim2022high,zhao2025microwave}, further validating that our model supporting CR drives can realize a wide variety of $n+m$ gates. Secondly, we validate the scalability of our approach by showing the results for five-qubit gates. Such as the CXXXX gate in Fig.~\ref{opt}(d). Fig.~\ref{opt}(d) illustrates the optimized pulse profiles, color coded for clarity. After 200 iterations, the fidelity of the CXXXX gate exceeds 92.1\%. Besides, we optimize the model for the CCXXX gate and achieve a fidelity of 92.6\%. These results confirm the applicability of our proposed model to multiqubit gates with higher qubit numbers.  For the five-qubit CXXXX gate, we additionally evaluate the error budget using a three-level description of each qubit, with a full Hilbert-space dimension of 243 and a computational-subspace dimension of 32. Leakage, coherence, and the remaining errors contribute 74.7\%,19.21\% and 6.09\% to the total infidelity, respectively. The remaining contribution, denoted by $\epsilon_{other}$, includes errors not assigned to leakage or coherence and is not assumed to be negligible. However, a natural trade-off arises between qubit count and gate fidelity. The scalability of the scheme is inherently limited by increasing control complexity and the emergence of higher-order interactions as the system size grows. This is reflected in the fidelity reduction from $>99.6\%$ (three qubits) to roughly $90\%$ (five qubits). \par
As the number of qubits increases, the gate fidelity gradually decreases, with fidelities above 99\% for three-qubit gates, approximately 90\%–99\% for four- and five-qubit gates. Additional numerical simulations of larger systems yield fidelities below 90\%, further illustrating the challenge of maintaining high fidelity as the system size increases. This degradation is mainly attributed to the increasing complexity of residual interactions in the full multilevel system. In particular, higher energy levels give rise to additional unwanted two-body and three-body entangling interactions, whose number and combined effect increase with the system size. For comparison, when the optimization is performed for an idealized system without these higher-level residual interactions, fidelities of approximately 99.99\% for three-qubit and five-qubit gates can still be achieved, indicating that the optimization procedure itself is not the dominant limitation. We further compare the performance of different optimization algorithms in Appendix E, as summarized in Table~\ref{com}, which shows that the choice of optimization algorithm has a comparatively smaller impact on the final fidelity. The increasing number of interaction channels may further contribute to the fidelity degradation at larger system sizes. While the framework can be extended to larger systems, maintaining high fidelity remains challenging because of increasingly complex residual interactions. Improved pulse parameterizations and optimization strategies may help mitigate these effects, although the present results do not identify the choice of optimization algorithm as the dominant limitation.\par

To benchmark the efficiency of our approach, we compare the performance of the proposed single-step gates with standard circuit decompositions. We distinguish here between the intrinsic error associated with the constituent two-qubit gates and the additional error expected from decoherence during the total execution time. We emphasize that the optimized fidelities reported in this section are obtained from coherent Hamiltonian simulations and do not include the \(T_1\) and \(T_2\) processes during the waveform optimization.

In the three-qubit regime, a CXX gate can be decomposed into two CNOT gates. Assuming a CNOT fidelity of \(99.9\%\), the fidelity associated with the two sequential CNOT gates is

$$
F_{\mathrm{gate}}^{\mathrm{CXX}}=(0.999)^2\approx99.8\%,
$$
corresponding to an intrinsic decomposition infidelity of approximately \(0.2\%\). This value is slightly lower than the coherent infidelity of our optimized single-step CXX gate, which is approximately \(0.4\%\). Therefore, the advantage of the single-step implementation should not be interpreted as an intrinsic-fidelity improvement over an idealized two-CNOT decomposition. Instead, its main advantage is the reduction in circuit depth and execution time. Assuming a duration of \(70\) ns for each CNOT gate, the decomposed implementation requires approximately \(140\) ns, whereas our single-step implementation is completed in approximately \(60\) ns.\par
The pulse optimization and the fidelities reported above are based on coherent Hamiltonian simulations without environmental relaxation or dephasing. To assess the potential benefit of the shorter execution time under realistic conditions, we now separately estimate the effects of decoherence for the two implementations. We use \(T_1=40~\mu\mathrm{s}\) and \(T_2=10~\mu\mathrm{s}\). Under an independent relaxation and dephasing model for the three qubits, the corresponding decoherence-only fidelity is approximately \(99.11\%\) for a \(60\)-ns evolution and \(97.93\%\) for a \(140\)-ns evolution. Therefore, although the idealized two-CNOT decomposition exhibits a slightly higher coherent fidelity, the shorter duration of the single-step implementation substantially reduces the exposure to relaxation and dephasing.

The difference in execution time becomes considerably more pronounced for the CCXXX gate. The decomposed implementation requires 18 two-qubit gates, as shown in Appendix~\ref{Appen_opt}. Assuming the same \(99.9\%\) fidelity for each two-qubit gate, the gate-only fidelity of the decomposition is

$$
F_{\mathrm{gate}}^{\mathrm{CCXXX}}=(0.999)^{18}\approx98.2\%,
$$

which is higher than the coherent fidelity of our optimized single-step CCXXX gate, \(92.6\%\). However, the 18-gate decomposition requires a total duration of approximately \(1260\) ns, compared with approximately \(60\) ns for the proposed single-step implementation. Using the same decoherence model, the decoherence-only fidelity decreases to approximately \(71.61\%\) over \(1260\) ns, whereas it remains approximately \(98.38\%\) over \(60\) ns. This comparison illustrates that the principal benefit of the direct multiqubit implementation arises from the substantial reduction in circuit depth and total execution time, rather than from a universal improvement in the intrinsic coherent fidelity.\par
We emphasize that the coherent gate fidelities and the decoherence-only fidelities are reported separately. We do not directly multiply the two fidelities or add the corresponding infidelities to obtain a total gate fidelity, since such a combination would require additional assumptions regarding the underlying error channels and their interplay during the driven dynamics.

A similar consideration applies to larger multiqubit subcircuits. Conventional decompositions generally require an increasing number of two-qubit operations as the number of qubits grows, leading to a simultaneous accumulation of gate errors and increased exposure to decoherence. In contrast, the proposed single-step approach avoids a long sequence of constituent two-qubit gates. The fidelity of the present five-qubit implementation remains limited by residual interactions in the multilevel system, including higher-order crosstalk. Recent developments in quantum optimal control~\cite{lewis2025quantumoptimalcontrolgeodesic} motivate exploring improved pulse parameterizations and optimization strategies as possible means of mitigating these effects. Nevertheless, our results for multiqubit gates with up to \(N=5\) demonstrate that direct single-step implementations can provide a substantial reduction in circuit depth and execution time, which is particularly advantageous for critical multiqubit subcircuits in the presence of decoherence.

While we assume a common frequency for target qubits to clarify the analytical derivation, this constraint can be relaxed in experimental implementations. By modulating multi-tone microwave drives to match distinct qubit transitions, the proposed scheme effectively bypasses frequency crowding while maintaining high fidelity within the standard fixed-frequency transmon architecture. We verify this through similar optimizations and achieve the same fidelity as before. The fidelity reaches 99.6\% and 90\% for the three- and five qubit models, which further verifies the advances of our regime after solving the frequency crowding issue. The results are shown in Appendix~\ref{Appen_opt}.

\begin{figure}[htp]
\centering
\includegraphics[width=8.7cm ]{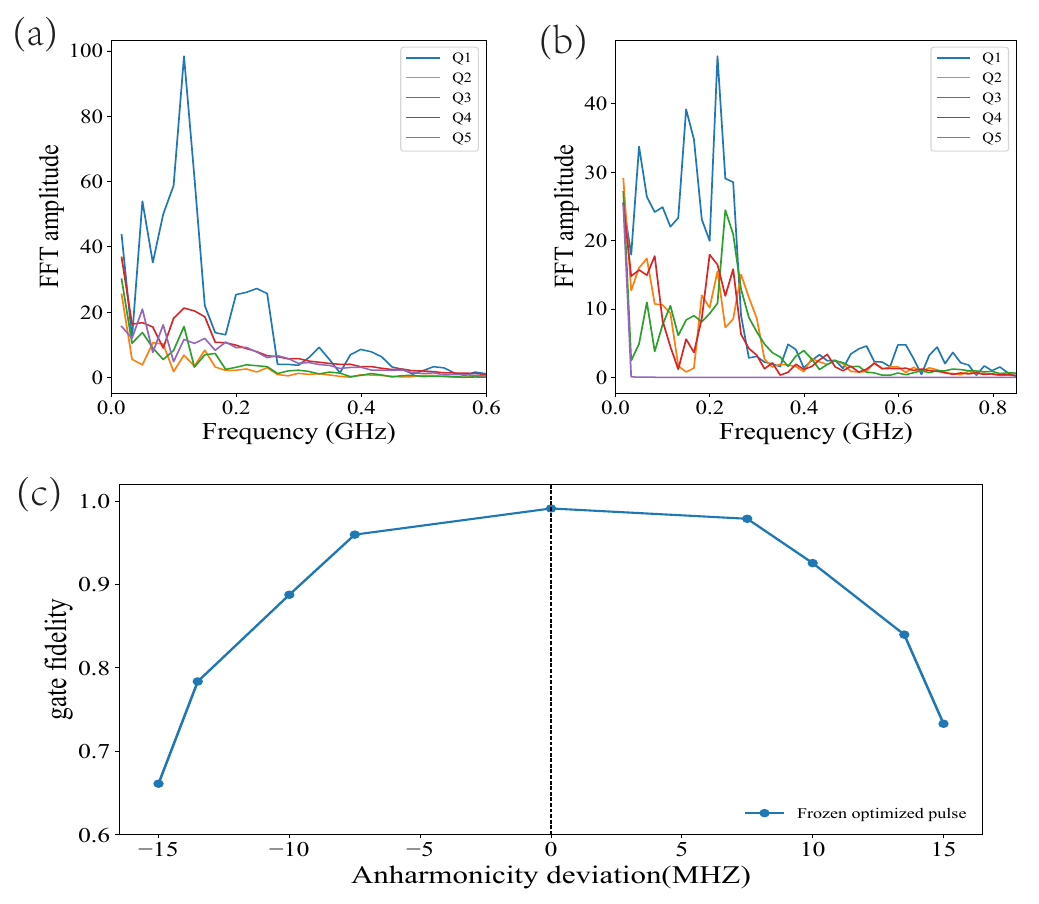}
\caption{Frequency spectra of the optimized control pulses for (a) the three-qubit \(CXX\) gate and (b) the five-qubit gate. (c) Sensitivity of the optimized
gates to anharmonicity variations }
\label{band}
\end{figure}
\section{experimental feasibility}
To assess the experimental realizability of the optimized control pulses, we further analyzed their frequency-domain characteristics. For a control envelope \(\Omega(t)\), its Fourier spectrum is defined as

$$ \tilde{\Omega}(f)=\int \Omega(t)e^{-i2\pi ft}\,dt, $$

and the corresponding spectral power is \(P(f)=|\tilde{\Omega}(f)|^2\). We characterize the effective bandwidth using the 90\% spectral-power bandwidth \(B_{90}\), defined by

$$ \frac{\displaystyle\int_{0}^{B_{90}}|\tilde{\Omega}(f)|^2df} {\displaystyle\int_{0}^{\infty}|\tilde{\Omega}(f)|^2df}=0.90. $$

As a representative example, we calculated the spectra of the optimized pulses for a three-qubit \(CXX\) gate and a five-qubit gate, as shown in Fig.~\ref{band}. For each gate, we calculate \(B_{90}\) separately for every optimized control pulse and take the largest value. This maximum is approximately 450 MHz for the three-qubit gate and 400 MHz for the five-qubit gate.Thus, although the optimized pulses exhibit nontrivial temporal structures, their dominant spectral components remain within the sub-GHz regime. These bandwidth requirements are compatible with experimentally demonstrated microwave control electronics for superconducting qubits. Superconducting-qubit experiments have demonstrated AWG-based control with bandwidths of several hundred MHz and sampling rates up to 2 GS/s~\cite{2010Quantum}. More recently, an integrated cryogenic RF AWG for transmon control achieved a 500 MHz signal bandwidth while maintaining qubit coherence comparable to conventional room-temperature control~\cite{Yves2017Low}. These results indicate that the bandwidth requirements of our optimized pulses are within the range of experimentally demonstrated microwave control technologies, supporting their practical implementability.\par

The anharmonicities adopted in this work are experimentally realistic for transmon qubits. Anharmonicities of a few hundred megahertz are routinely achievable; for example, transmon devices have been designed with ($\alpha/2\pi=300$) MHz and ($\alpha/2\pi\approx-346$) MHz, while experimentally measured values around (-398) MHz have also been reported~\cite{2015Building,2020Breaking}. Therefore, the values of (-300), (-320), and (-350) MHz used in our three-qubit simulations, as well as (-300) MHz used for the five-qubit simulations, are within experimentally accessible ranges.\par
We further investigate the sensitivity of the optimized gates to anharmonicity variations by varying the anharmonicities around their nominal values while keeping the optimized waveforms unchanged in Fig.~\ref{band}(c). The fidelity degradation under such variations is expected because the control pulses are optimized for a specific Hamiltonian spectrum, and deviations in anharmonicity modify the higher-level transition frequencies and effective CR interactions. Nevertheless, the proposed gate scheme does not rely on a fixed anharmonicity configuration. These results highlight the sensitivity of the optimized waveforms to anharmonicity variations and the importance of calibrating the control pulses to the specific device parameters.\par
For the other parameters in our systems, we further verify the validity of the RWA used in deriving the effective Hamiltonian. The validity of the RWA is ensured by the small ratios $\Omega_{\max}/(2\omega_t)\approx9.6\times10^{-3}$
and $ g/(2\omega_t)\approx7.7\times10^{-4}. $
Moreover, the neglected counter-rotating terms oscillate more than 600 cycles during the 60-ns gate operation, leading to their effective averaging.These parameters are also within the experimentally demonstrated regime of superconducting transmon systems~\cite{Arute2019Quantum}.

\section{Conclusion}
In this paper, we present a comprehensive framework for constructing $(n+m)$ multiqubit gates using CR driving on transmon networks. This architecture offers a practical pathway for designing next-generation quantum devices and optimizing algorithmic circuit depths. We introduce the models for these multiqubit gates and two methods for computing the effective interactions within our model via quad frame rotation and block diagonalization. Finally, we present the results of pulse optimization for our multiqubit gates using a gradient-based algorithm, demonstrating achievable fidelities that affirm the experimental viability of the proposed gates.
Compared with existing CR-based multiqubit gate proposals,
our work connects conditional gate construction with multilevel
interaction analysis for representative control--target configurations.
As summarized in Table~\ref{tab:comparison_cr}, the proposed framework
combines parallel CR-induced two-body $ZX$ interactions with
independently adjustable local $X/Y$ terms to construct a family
of controlled-Pauli gates. We derive conditions for the required
target evolution in different control-state sectors and analyze
parasitic interactions arising in the multilevel transmon model.
The resulting interaction coefficients and their parameter dependence
guide parameter selection and subsequent pulse optimization.

\begin{table*}[t]
\centering
\caption{Comparison of representative CR-based multiqubit gate
approaches and the present work. Entries summarize the scope
investigated in each work.}
\label{tab:comparison_cr}

\small
\renewcommand{\arraystretch}{1.25}
\setlength{\tabcolsep}{4pt}
\setlength{\arrayrulewidth}{0.4pt}

\begin{tabularx}{\textwidth}{
    >{\raggedright\arraybackslash\bfseries}p{1.9cm}
    *{5}{>{\raggedright\arraybackslash}X}
}
\hline
\textbf{Comparison}
&
\textbf{Kim \textit{et al.}}~\cite{kim2022high}
&
\textbf{Gokhale \textit{et al.}}~\cite{2020Quantum}
&
\textbf{Itoko \textit{et al.}}~\cite{2023Three}
&
\textbf{Xu \textit{et al.}}~\cite{xu2025parity}
&
\textbf{This work}
\\
\hline

Gate scope
&
$i$Toffoli and other three-qubit gates
&
Multi-target fan-out
&
Three-qubit parity gate
&
Three-qubit parity gate
&
Controlled-Pauli gates in multiple control--target configurations
\\
\hline

Interaction resource
&
Two-body CR
&
Parallel CR
&
Parallel two-body CR
&
Three-body $ZZX$
&
Parallel two-body $ZX$ with local $X/Y$ drives
\\
\hline

Design focus
&
Conditional target flipping
&
Simultaneous multi-target flipping
&
Simultaneous parity mapping
&
Three-body interaction engineering
&
Conditional gate construction and parasitic-interaction analysis
\\
\hline

Theoretical treatment
&
QFR effective model
&
Parallel fan-out construction
&
Schrieffer--Wolff expansion
&
Perturbative analysis and BD
&
QFR, numerical BD, and BB expansion
\\
\hline

Validation
&
Experiment and simulation
&
Experimental proof of concept
&
Experiment
&
Simulation
&
Analytical and numerical; fidelity trends for 3--6 qubits
\\
\hline
\end{tabularx}

\end{table*}

The utilization of our multiqubit gates in quantum circuits offers significant advantages in reducing circuit depth ~\cite{chu2023scalable} compared with decomposed implementations. Our proposed $(n+m)$ gates closely resemble stabilizers in quantum error correction codes, and we are actively exploring additional quantum algorithms that could leverage these multiqubit gates as efficient computational resources. Moreover, building on the theoretical feasibility discussed above, we aim to develop and validate multiqubit gates with arbitrary target rotations beyond X, Y, and Z operations in future work. However, a natural trade-off exists between gate fidelity and the number of qubits involved due to the increasing complexity of higher-order interactions. To address this, we are investigating strategies to mitigate such a trade-off, for instance, we plan to explore whether hyperbolic pulses can help improve the fidelity of multiqubit gates in larger systems~\cite{economou2015analytical,barnes2012analytically}, which allow precise analytical control over the system dynamics without relying on approximations. Additionally, rather than merely suppressing three-body interactions (e.g., $ZXX$), future work will explore harnessing these terms as unique resources for synthesizing novel multiqubit gate dynamics~\cite{xu2025parity}. Furthermore, the CR regime may also be applicable to other quantum platforms, such as quantum dots, for the construction of multiqubit gates, and we intend to pursue this direction in future research. This work serves as a foundation for studying multi-interaction systems and multiqubit gates, as well as an effective methodology for quantum circuit design.

\section*{Author Contributions}

X.-H.D. conceived and supervised the project. G.W., Y.S., and X.-H.D. developed the theoretical framework and designed the gate architecture. G.W. developed the analytical theory, carried out the analytical derivations, and performed the numerical simulations and data analysis. X.-H.D. and Y.S. supervised the theoretical development and numerical work. G.W. and X.-H.D. wrote the manuscript. All authors contributed to the manuscript.

\section*{Data and code availability}
Core code, data, interval certificates, and figure sources will be made
available at manuscript submission in
\href{https://github.com/QDynamics/CalibrationCostOfQControl}
{\texttt{QDynamics/CalibrationCostOfQControl}}.

\begin{acknowledgments}
This work was supported by the Key-Area Research and Development Program of Guang-Dong Province (Grant No. 2018B030326001), Innovation Program for Quantum Science and Technology (2024ZD0300400), the National Natural Science Foundation of China (U1801661), the Guangdong Provincial Key Laboratory (Grant No.2019B121203002), and the Science, Technology and Innovation Commission of Shenzhen Municipality (JCYJ20170412152620376, KYTDPT20181011104202253), Shenzhen Science and Technology Program (KQTD20200820113010023), and the National Natural Science Foundation of China (Grant No 12404566)  
\end{acknowledgments}

\appendix
\section{Deriving effective Hamiltonian by a way of quad frame rotation}
\label{Appen_Heff1}
\subsection{effective Hamiltonian for one control and two target model}
In this model, we propose the design of an CXX (or CZX) gate, with the schematic representation detailed in Fig.~\ref{model001}(b). 

Accordingly, the initial Hamiltonian can be formulated as:
\begin{equation}\begin{aligned}
H=&\frac{\omega_t}{2}\sigma_{t1}^z+\frac{\omega_t}{2}\sigma_{t2}^z+\frac{\omega_c}{2}\sigma_c^z+\frac{g_1}{2}\sigma_{t1}^x\sigma_{c}^x+\frac{g_2}{2}
\sigma_c^x\sigma_{t2}^x\\
&+\Omega_{t1}\text{cos}(\omega_tt+\phi_{t1})\sigma_{t1}^x+\Omega_{t2}\text{cos}(\omega_tt+\phi_{t2}^x)\sigma_{t2}^x\\&+\Omega_c\text{cos}(\omega_tt+\phi_c)\sigma_c^x\end{aligned}\label{r1}\end{equation}
the initial Hamiltonian can be transformed into a rotating frame through the application of an appropriate unitary transformation $H'=U^{-1}HU-iU^{-1}\frac{dU}{dt}$ with $U_a=\text{e}^{-\frac{it}{2}(\omega_t\sigma_{t1}^z+\omega_t\sigma_{t2}^z+\omega_t\sigma_c^z)}$,
 \begin{equation}\begin{aligned}
H_a=&\frac{\delta_c}{2}\sigma_{c}^z +\frac{g_1}{2}[\text{cos}(\omega_tt)\sigma_{t1}^x-\text{sin}(\omega_tt)\sigma_{t1}^y][\text{cos}(\omega_tt)\sigma_{c}^x\\&-\text{sin}(\omega_tt)\sigma_{c}^y]+\frac{g_2}{2}
[\text{cos}(\omega_tt)\sigma_{c}^x-\text{sin}(\omega_tt)\sigma_{c}^y]\\&[\text{cos}(\omega_tt)\sigma_{t2}^x-\text{sin}(\omega_tt)\sigma_{t2}^y]+\Omega_{t1}\text{cos}(\omega_tt+\phi_{t1})\\&[\text{cos}(\omega_tt)\sigma_{t1}^x-\text{sin}(\omega_tt)\sigma_{t1}^y]+\Omega_{t2}\text{cos}(\omega_tt+\phi_{t2}^x)\\&[\text{cos}(\omega_tt)\sigma_{t2}^x-\text{sin}(\omega_tt)\sigma_{t2}^y]+\Omega_c\text{cos}(\omega_tt+\phi_c)\\&[\text{cos}(\omega_tt)\sigma_{c}^x-\text{sin}(\omega_tt)\sigma_{c}^y]\end{aligned}\end{equation}
then we rotate it into the double rotating frame with $U_b=\text{e}^{-\frac{i}{2}\phi_c\sigma_c^z}$,
 \begin{equation}\begin{aligned}
H_b=&\frac{\delta_c}{2}\sigma_{c}^z +\frac{g_1}{2}[\text{cos}(\omega_tt)\sigma_{t1}^x-\text{sin}(\omega_tt)\sigma_{t1}^y]\text{cos}(\omega_tt+\phi_c)\\&\sigma_{c}^x-\text{sin}(\omega_tt+\phi_c)\sigma_{c}^y]+\frac{g_2}{2}
[\text{cos}(\omega_tt+\phi_c)\sigma_{c}^x-\text{sin}\\&(\omega_tt+\phi_c)\sigma_{c}^y][\text{cos}(\omega_tt)\sigma_{t2}^x-\text{sin}(\omega_tt)\sigma_{t2}^y]+\Omega_{t1}\text{cos}\\&(\omega_tt+\phi_{t1})[\text{cos}(\omega_tt)\sigma_{t1}^x-\text{sin}(\omega_tt)\sigma_{t1}^y]+\Omega_{t2}\text{cos}(\omega_tt\\&+\phi_{t2}^x)[\text{cos}(\omega_tt)\sigma_{t2}^x-\text{sin}(\omega_tt)\sigma_{t2}^y]+\Omega_c\text{cos}(\omega_tt+\phi_c\\&)[\text{cos}(\omega_tt+\phi_c)\sigma_{c}^x-\text{sin}(\omega_tt+\phi_c)\sigma_{c}^y]\end{aligned}\end{equation}
the trigonometric identity $\text{cos}(\alpha)\text{cos}(\beta)=\frac{1}{2}(\text{cos}(\alpha+\beta)+\text{cos}(\alpha-\beta)$ may be employed in this context. Utilizing the RWA, all terms oscillating at $2\omega_t$ can be neglected. As a result, $H_b$ reduces to an effective Hamiltonian in the double rotating frame,
 \begin{equation}\begin{aligned}H_{b}^{eff}=&\frac{\delta_c}{2}\sigma_c^z+\frac{\Omega_c}{2}\sigma_c^x+\frac{\Omega_{t1}}{2}\text{cos}\phi_{t1}\sigma_{t1}^x+\frac{\Omega_{t1}}{2}\text{sin}\phi_{t1}\sigma_{t1}^y\\&+\frac{\Omega_{t2}}{2}\text{cos}\phi_{t2}\sigma_{t2}^x+\frac{\Omega_{t2}}{2}\text{sin}\phi_{t2}\sigma_{t2}^y+\frac{g_1}{2}[\text{cos}(\omega_tt\\&+\phi_c)\sigma_c^x-\text{sin}(\omega_tt+\phi_c)\sigma_c^y](\text{cos}(\omega_tt)\sigma_{t1}^x-\text{sin}\\&(\omega_tt)\sigma_{t1}^y)+\frac{g_2}{2}[\text{cos}(\omega_tt+\phi_c)\sigma_c^x-\text{sin}(\omega_tt+\phi_c)\\&\sigma_c^y](\text{cos}(\omega_tt)\sigma_{t2}^x-\text{sin}(\omega_tt)\sigma_{t2}^y)\end{aligned}\end{equation}
then another rotation is applied,$U_c=\text{e}^{-\frac{i}{2}\xi_c\sigma_c^y}$ and $\xi_c=tan^{-1}(-\delta_c/\Omega_c)$,
\begin{equation}\begin{aligned}H_{c}=&\frac{\delta_c}{2}(-\text{sin}(\xi_c)\sigma_c^x+cos(\xi_c)\sigma_c^z)+\frac{\Omega_c}{2}(\text{cos}(\xi_c)\sigma_c^x\\&+\text{sin}(\xi_c)\sigma_c^z)+\frac{\Omega_{t1}}{2}\text{cos}\phi_{t1}\sigma_{t1}^x+\frac{\Omega_{t1}}{2}\text{sin}\phi_{t1}\sigma_{t1}^y\\&+\frac{\Omega_{t2}}{2}\text{cos}\phi_{t2}\sigma_{t2}^x+\frac{\Omega_{t2}}{2}\text{sin}\phi_{t2}\sigma_{t2}^y+\frac{g_1}{2}[\text{cos}(\omega_tt\\&+\phi_c)(\text{cos}(\xi_c)\sigma_c^x+\text{sin}(\xi_c)\sigma_c^z)-\text{sin}(\omega_tt+\phi_c)\\&\sigma_c^y](\text{cos}(\omega_tt)\sigma_{t1}^x-\text{sin}(\omega_tt)\sigma_{t1}^y)+\frac{g_2}{2}[\text{cos}(\omega_tt\\&+\phi_c)(\text{cos}(\xi_c)\sigma_c^x+\text{sin}(\xi_c)\sigma_c^z)-\text{sin}(\omega_tt+\phi_c\\&)\sigma_c^y](\text{cos}(\omega_tt)\sigma_{t2}^x-\text{sin}(\omega_tt)\sigma_{t2}^y)\end{aligned}\end{equation}
certain terms in this equation can be compensated through a fourth rotation, defined by the unitary operator $U_d=\text{e}^{-\frac{i}{2}\eta_ct\sigma_c^x}$,$\eta_c=\sqrt{\delta^2_c+\Omega^2_c}$. This transformation rotates the Hamiltonian into a quadruple rotating frame.
\begin{equation}\begin{aligned}H_{d}=&\frac{\delta_c}{2}(-\text{sin}(\xi_c)\sigma_c^x+\text{cos}(\xi_c)\sigma_c^z)+\frac{\Omega_c}{2}(\text{cos}(\xi_c)\sigma_c^x\\&+\text{sin}(\xi_c)\sigma_c^z)+\frac{\Omega_{t1}}{2}\text{cos}\phi_{t1}\sigma_{t1}^x+\frac{\Omega_{t1}}{2}\text{sin}\phi_{t1}\sigma_{t1}^y\\&+\frac{\Omega_{t2}}{2}\text{cos}\phi_{t2}\sigma_{t2}^x+\frac{\Omega_{t2}}{2}\text{sin}\phi_{t2}\sigma_{t2}^y+\frac{g_1}{2}[\text{cos}(\omega_tt\\&+\phi_c)(\text{cos}(\xi_c)\sigma_c^x+\text{sin}(\xi_c)(\text{cos}(\eta_ct)\sigma_c^z+\text{sin}(\\&\eta_ct)\sigma_c^y)-\text{sin}(\omega_tt+\phi_c)(-\text{sin}(\eta_ct)\sigma_c^z+\text{cos}(\eta_ct\\&)\sigma_c^y)](\text{cos}(\omega_tt)\sigma_{t1}^x-\text{sin}(\omega_tt)\sigma_{t1}^y)+\frac{g_2}{2}[\text{cos}(\omega_tt\\&+\phi_c)(\text{cos}(\xi_c)\sigma_c^x+\text{sin}(\xi_c) (\text{cos}(\eta_ct)\sigma_c^z+\text{sin}\\&(\eta_ct)\sigma_c^y))-\text{sin}(\omega_tt+\phi_c)(-\text{sin}(\eta_ct)\sigma_c^z+\text{cos}\\&(\eta_ct)\sigma_c^y)](\text{cos}(\omega_tt)\sigma_{t2}^x-\text{sin}(\omega_tt)\sigma_{t2}^y)\end{aligned}\end{equation}
all terms oscillating at very high frequencies (including $\eta_ct$ and $2\omega_tt$) are eliminated through the RWA, yielding an effective Hamiltonian in the quadruple rotating frame,\begin{equation}\begin{aligned}H_{d}^{eff}=&\frac{\Omega_{t1}}{2}\text{cos}\phi_{t1}\sigma_{t1}^x+\frac{\Omega_{t1}}{2}\text{sin}\phi_{t1}\sigma_{t1}^y+\frac{\Omega_{t2}}{2}\text{cos}\phi_{t2}\sigma_{t2}^x\\&+\frac{\Omega_{t2}}{2}\text{sin}\phi_{t2}\sigma_{t2}^y+\frac{g_1}{4}\text{cos}\xi_c(\text{cos}\phi_c\sigma_c^x\sigma_{t1}^x+\text{sin}\phi_c\\&\sigma_c^x\sigma_{t1}^y)+\frac{g_2}{4}\text{cos}\xi_c(\text{cos}\phi_c\sigma_c^x\sigma_{t2}^x+\text{sin}\phi_c\sigma_c^x\sigma_{t2}^y)\end{aligned}\end{equation}
finally, we apply a final transformation through the unitary operator $U_e=\text{e}^{-\frac{i}{2}\xi_c\sigma_c^y}$. Due to the condition $\Omega_c<<\delta_c$, which is typically adopted in experimental implementations, the XX and XY interaction terms in the Hamiltonian can be disregarded as their contributions are substantially smaller than those of the ZX terms. Moreover, by selecting $\phi_c=0$, all ZY interactions are effectively removed. As a result, the effective Hamiltonian of the system is derived as follows.
\begin{equation}\begin{aligned}
H^{eff}&=\frac{\Omega_{t1}}{2}\text{cos}\phi_{t1}\sigma_{t1}^x+\frac{\Omega_{t1}}{2}\text{sin}\phi_{t1}\sigma_{t1}^y+\frac{\Omega_{t2}}{2}\text{cos}\phi_{t2}\sigma_{t2}^x\\&+\frac{\Omega_{t2}}{2}\text{sin}\phi_{t2}\sigma_{t2}^y+\frac{g_1\Omega_c}{4\delta_c}\sigma_c^z\sigma_{t1}^x+\frac{g_2\Omega_c}{4\delta_c}\sigma_c^z\sigma^x_{t2}
\end{aligned}\label{r2}\end{equation}\par
Moreover, if two additional pulses resonant with each target qubit are applied, the total number of pulses on the two target qubits amounts to four, $\Omega_{t1}\text{cos}(\omega_tt+\phi_{t1})$,$\Omega_{t2}\text{cos}(\omega_tt+\phi_{t2})$ and $\Omega_{t1}'\text{cos}(\omega_tt+\phi_{t1}')$,$\Omega_{t2}'\text{cos}(\omega_tt+\phi_{t2}^{'})$,so the initial Hamiltonian is \begin{equation}\begin{aligned}
H=&\frac{\omega_t}{2}\sigma_{t1}^z+\frac{\omega_t}{2}\sigma_{t2}^z+\frac{\omega_c}{2}\sigma_c^z+\frac{g_1}{2}\sigma_{t1}^x\sigma_{c}^x+\frac{g_2}{2}
\sigma_c^x\sigma_{t2}^x\\&+\Omega_{t1}\text{cos}(\omega_tt+\phi_{t1})\sigma_{t1}^x+\Omega_{t2}\text{cos}(\omega_tt+\phi_{t2})\sigma_{t2}^x\\&+\Omega_c\text{cos}(\omega_tt+\phi_c)\sigma_c^x+\Omega_{t1}^{'} \text{cos}(\omega_tt+\phi_{t1}^{'})\sigma_{t1}^x\\&+\Omega_{t2}^{'}\text{cos}(\omega_tt+\phi_{t2}^{'})\sigma_{t2}^x\end{aligned}\end{equation}the computation of these two terms is identical to that of the local drives described above. Consequently, the effective Hamiltonian following calculation becomes
\begin{equation}\begin{aligned}
H^{eff}&=\frac{\Omega_{t1}}{2}\text{cos}\phi_{t1}\sigma_{t1}^x+\frac{\Omega_{t1}}{2}]text{sin}\phi_{t1}\sigma_{t1}^y+\frac{\Omega_{t2}}{2}\text{cos}\phi_{t2}\sigma_{t2}^x\\&+\frac{\Omega_{t2}}{2}\text{sin}\phi_{t2}\sigma_{t2}^y+\frac{g_2\Omega_c}{4\delta_c}\sigma_c^z\sigma_{t2}^x+\frac{g_1\Omega_c}{4\delta_c}\sigma_c^z\sigma^x_{t1}+\\&\frac{\Omega_{t1}^{'}}{2}\text{cos}\phi_{t1}^{'}\sigma_{t1}^x+\frac{\Omega_{t1}^{'}}{2}\text{sin}\phi_{t1}^{'}\sigma_{t1}^y+\frac{\Omega_{t2}^{'}}{2}\text{cos}\phi_{t2}^{'}\sigma_{t2}^x+\\&\frac{\Omega_{t2}^{'}}{2}\text{sin}\phi_{t2}^{'}\sigma_{t2}^y
\end{aligned}\end{equation}\par
For the parameters considered in this work, the validity of the approximations
employed in the quad-frame rotation can be quantitatively assessed. For the parameters used in this work, the validity of the approximations can
be quantitatively assessed. After the second unitary transformation, the
neglected counter-rotating terms are suppressed by
\[
\frac{\Omega_c}{2\omega_t}=4.8\times10^{-3},
\qquad
\frac{g}{2\omega_t}=7.7\times10^{-4}.
\]
After the fourth transformation, $\eta_c=\sqrt{\Omega_c^2+\delta_c^2}$ gives
$\eta_c/2\pi\approx304.1$ MHz, and the higher-order correction satisfies
\[
\frac{g}{\eta_c}=2.6\times10^{-2}.
\]
Meanwhile, the residual $XX$ and $YY$ terms are suppressed in the dispersive
regime with $\Omega_c/\delta_c=0.167$. These small corrections validate the
approximations adopted in deriving the effective Hamiltonian.

\subsection{effective Hamiltonian for two control and one target model}
In this model, we present the implementation of a two control one target  gate shown in Fig.~\ref{model001}(c), specifically the Toffoli gate (CCNOT). 

The initial Hamiltonian is expressed as follows:
\begin{equation}\begin{aligned}
H=&\frac{\omega_t}{2}\sigma_t^{z}+\frac{\omega_{c1}}{2}\sigma_{c1}^{z}+\frac{\omega_{c2}}{2}\sigma_{c2}^{z}+\Omega_{c1}\text{cos}(\omega_tt+\phi_{c1})\sigma_{c1}^x\\&+\Omega_{c2}\text{cos}(\omega_tt+\phi_{c2})\sigma_{c2}^x+\Omega_{t}\text{cos}(\omega_tt+\phi_{t})\sigma_{t}^x+\\&\frac{g_1}{2}\sigma_{t}^x\sigma_{c1}^x+\frac{g_2}{2}\sigma_{t}^x\sigma_{c2}^x
\end{aligned}\label{r3}\end{equation}with two rotation transformation $U_a=\text{e}^{-\frac{it}{2}(\omega_t\sigma_{t1}^z+\omega_t\sigma_{t2}^z+\omega_t\sigma_c^z)}$,$U_b=\text{e}^{-\frac{i}{2}\phi_{c1}\sigma_{c1}^z-\frac{i}{2}\phi_{c2}\sigma_{c2}^z}$ and RWA the effective Hamiltonan in double rotating frame is,\begin{equation}\begin{aligned}
H_{df}=&\frac{\delta_{c1}}{2}\sigma_{c1}^z+\frac{\delta_{c2}}{2}\sigma_{c2}^z+\frac{g_1}{2}[(-\text{sin}(\varphi_{c1})\sigma_{c1}^x-\text{cos}(\varphi_{c1})\\&\sigma_{c1}^y)sin(\omega_tt)+\text{cos}(\omega_tt)(-\sigma_{c1}^y\text{sin}(\varphi_{c1})+\sigma_{c1}^x\\&\text{cos}(\varphi_{c1}))](-\sigma_t^y\text{sin}(\omega_tt)+\sigma_t^x\text{cos}(\omega_t^x))+\frac{g_2}{2}\\&(-\sigma_t^y\text{sin}(\omega_tt)+\sigma_t^x\text{cos}(\omega_tt))[(-\sigma_{c2}^x\text{sin}(\varphi_{c2})\\&+\sigma_{c2}^y\text{cos}(\varphi_{c2}))\text{sin}(\omega_tt)+\text{cos}(\omega_tt)(-\sigma_{c2}^y\text{sin}(\varphi_{c2})\\&+\sigma_{c2}^x\text{cos}(\varphi_{c2}))]+\Omega_t\text{cos}(\omega_tt+\varphi_{t})(-\sigma_t^y\text{sin}(\omega_tt)\\&+\sigma_t^x\text{cos}(\omega_tt))+\Omega_{c1}\text{cos}(\omega_tt+\varphi_{c1})[\text{cos}(\omega_tt+\\&\varphi_{c1})\sigma_{c1}^x-\text{sin}(\omega_tt+\varphi_{c1})\sigma_{c1}^y]+\Omega_{c2}\text{cos}(\omega_tt+\varphi_{c2})\\&[\text{cos}(\omega_tt+\varphi_{c2})\sigma_{c2}^x-\text{sin}(\omega_tt+\varphi_{c2})\sigma_{c2}^y]
\end{aligned}\end{equation}
after two more rotation transformation $U_c=\text{e}^{-\frac{i}{2}\xi_{c1}\sigma_{c1}^y-\frac{i}{2}\xi_{c2}\sigma_{c2}^y}$ and $U_d=\text{e}^{-\frac{i}{2}\eta_{c1}\sigma_{c1}^x-\frac{i}{2}\eta_{c2}\sigma_{c2}^x}$ ,
 \begin{equation}\begin{aligned}
H_{qf}=&\frac{\Omega_t}{2}\sigma_t^y\sin(\varphi_t)+\frac{\Omega_t}{2}\sigma_t^x\cos(\varphi_t)+\frac{g_1}{2}[\cos(\omega_tt+\varphi_{c1})\\&(\sigma_{c1}^x\cos(\xi_{c1})+\sin(\xi_{c2})(\sigma_{c1}^y\sin(\eta_{c1}t)+\sigma_{c1}^z\cos(\eta_{c1}t))\\&-\sin(\omega_tt+\varphi_{c1})(\sigma_{c1}^z\cos(\eta_{c1}t)+\sigma_{c1}^y\sin(\eta_{c1}t))] \\&(-\sigma_t^y\sin(\omega_tt)+\sigma_t^x\cos(\omega_{t}t))+\frac{g_2}{2}(-\sigma_t^y\sin(\omega_tt)\\&+\sigma_t^x\cos(\omega_tt)) [\cos(\omega_tt+\varphi_{c2})(\sigma_{c2}^z\cos(\xi_{c1})\\&+\sin(\xi_{c1})(\sigma_{c2}^y\sin(\eta_{c1}t)+\sigma_{c2}^z\cos(\eta_{c1}t))\\&-\sin(\omega_tt+\varphi_{c2})(\sigma_{c2}^z\cos(\eta_{c2}t)+\sigma_{c2}^y\cos(\eta_{c2}t))]
\end{aligned}\end{equation}
and RWA is applied,
 \begin{equation}\begin{aligned}
H_{qf}^{eff}=&\frac{\Omega_t}{2}\sigma_t^y\text{sin}(\varphi_t)+\frac{\Omega_t}{2}\sigma_t^x\text{cos}(\varphi_t)+\frac{g_1}{2}(\sigma_{c1}^x\sigma_t^y\frac{\text{sin}(\varphi_{c1})}{2}\\&+\sigma_{c1}^x\sigma_t^x\frac{\text{cos}(\varphi_{c1})}{2})\text{cos}(\xi_{c1})+\frac{g_2}{2}\text{cos}(\xi_{c2})(\sigma_t^y\sigma_{c2}^x\\&\frac{\text{sin}(\varphi_{c2})}{2}+\sigma_t^x\sigma_{c2}^x\frac{\text{cos}(\varphi_{c2})}{2})
 \end{aligned}\end{equation}
and with another rotation transformation, \begin{equation}\begin{aligned}
H_{qf}^{'eff}=&\frac{\Omega_t}{2}\sigma_t^y\text{sin}(\varphi_t)+\frac{\Omega_t}{2}\sigma_t^x\text{cos}(\varphi_t)+\frac{g_1}{4}[\text{sin}(\varphi_{c1})\\&(\sigma_{c1}^z\sigma_t^y\frac{\Omega_{c1}\delta_{c1}}{\eta_{c1}^2}+\sigma_{c1}^x\frac{\Omega_{c1}^2}{\eta_{c1}^2})+\text{cos}(\varphi_{c1})(\frac{\delta_{c1}\Omega_{c1}}{\eta_{c1}^2}\\&\sigma_{c1}^z\sigma_t^x+\sigma_{c1}^x\sigma_t^x\frac{\Omega_{c1}^2}{\eta_{c1}^2})]+\frac{g_2}{4}[\text{sin}(\varphi_{c2})(\sigma_t^y\sigma_{c2}^z\\&\frac{\Omega_{c2}\delta_{c2}}{\eta_{c2}^2}+\sigma_t^y\sigma_{c2}^x\frac{\Omega_{c2}^2}{\eta_{c2}^2})+cos(\varphi_{c2})(\sigma_t^x\sigma_{c2}^z\\&\frac{\delta_{c2}\Omega_{c2}}{\eta_{c2}^2}+\sigma_t^x\sigma_{c2}^x\frac{\Omega_{c2}^2}{\eta_{c2}^2})\end{aligned}\end{equation}
ignore those small terms as $\Omega<<\delta$,we get the effective Hamiltonian,
\begin{equation}\begin{aligned}
H^{eff}=&\frac{\Omega_t}{2}\sigma_t^y\text{sin}(\varphi_t)+\frac{\Omega_t}{2}\sigma_t^x\text{cos}(\varphi_t)+\frac{g_1\Omega_{c1}}{4\delta_{c1}}\\&(\sigma_{c1}^z\sigma_t^y\text{sin}(\varphi_{c1})+\sigma_{c1}^z\sigma_t^x\text{cos}(\varphi_{c1}))\\&+\frac{g_2\Omega_{c2}}{4\delta_{c2}}(\sigma_t^y\sigma_{c2}^z\text{sin}(\varphi_{c2})+\sigma_t^x\sigma_{c2}^z\text{cos}(\varphi_{c2}))
\end{aligned}\label{r4}\end{equation}

\section{Realization of multiqubit gates with the effective Hamiltonian}
\label{Appen_Heff}
The dynamic of our three qubit system is governed by the reduced effective Hamiltonian \begin{equation}\begin{aligned}
&H=\\&\alpha X_{t1}Z_{c}+\beta Z_{c}X_{t2}+\delta_{t1}Y_{t1}+\gamma_{t1}X_{t1}+\delta_{t2}Y_{t2}+\gamma_{t2}X_{t2}
\end{aligned}\label{r5}\end{equation}\par
To analyze the dynamics of the designed system, we project the Hamiltonian onto the target states using the reduction formula $H'= _c\bra{i}H\ket{i}_c$ , where $\ket{i}_c$ denotes the state of the control qubit.
\begin{equation}\begin{aligned}
H^0=\alpha X_{t1}+\beta X_{t2}+\delta_{t1}Y_{t1}+\gamma_{t1}X_{t1}+\delta_{t2}Y_{t2}+\gamma_{t2}X_{t2}\\
H^1=-\alpha X_{t1}-\beta X_{t2}+\delta_{t1}Y_{t1}+\gamma_{t1}X_{t1}+\delta_{t2}Y_{t2}+\gamma_{t2}X_{t2}
\end{aligned}\end{equation} resulting in a Hamiltonian that simplifies to a linear summation over the target qubits, allow us to analyze target qubits individually. Assuming the parameters $\alpha=\gamma_{t1}$ and $\beta=\gamma_{t2}$, we obtain $H^0=2\alpha X+\delta_{t1} Y$ for the control state $\ket{0}$ and $H^1=\delta_{t1} Y$ for the control state $\ket{1}$ when examining the target qubits separately. Since the effective Hamiltonian of the system is time-independent, the dynamics are governed by the unitary operator $U=\text{e}^{-i H t}$ , which determines the evolution of the system state.
\begin{equation}\begin{aligned}&\ket{\phi(t)}=\\&\text{cos}(\sqrt{4\alpha^2+\delta_{t1}^2}t)\ket{0}+\frac{-2i\alpha+\delta_{t1}}{\sqrt{4\alpha^2+\delta_{t1}^2}}\text{sin}(\sqrt{4\alpha^2+\delta_{t1}^2}t)\ket{1} \\&
\ket{\phi(t)}=\\&\frac{-2i\alpha-\delta_{t1}} {\sqrt{4\alpha^2+\delta_{t1}^2}}\text{sin}(\sqrt{4\alpha^2+\delta_{t1}^2}t)\ket{0}+\text{cos}(\sqrt{4\alpha^2+\delta_{t1}^2}t)\ket{1}\end{aligned}\end{equation} when the target qubit is initialized in state $\ket{0}$ with the control state $\ket{0}$ and when the target starts from $\ket{1}$ with the control state $\ket{0}$, the system evolves under the time condition $\tau=\frac{2\pi}{\sqrt{4\alpha^2+\delta_{t1}^2}}$. Upon satisfying this condition, the system returns to its initial state after evolution, making the target qubit remains unchanged when the control is $\ket{0}$. When control becomes $\ket{1}$: \begin{equation}\begin{aligned}&\ket{\phi(t)}=\text{cos}(\delta_{t1}t)\ket{0}+
\text{sin}(\delta_{t1}t)\ket{1}
\\&\ket{\phi(t)}=-\text{sin}(\delta_{t1}t)\ket{0}+\text{cos}(\delta_{t1}t)\ket{1}\label{aa}\end{aligned}\end{equation} the first solution corresponds to the initial state $\ket{0}$, while the second solution represents initial state is $\ket{1}$. If a X(Y) rotation operation is desired for target 1, the parameter $\delta_{t1}=\frac{\pi}{2\tau}$ can be adopted, resulting in $\ket{\phi(t)}=\ket{1}$ when the initial state is $\ket{0}$, and $\ket{\phi(t)}=-\ket{0}$ when the initial state is $\ket{1}$. Alternatively, if a Z operation is required, the parameter $\delta_{t1}=\frac{\pi}{\tau}$ may be employed, yielding $\ket{\phi(t)}=-\ket{0}$ for an initial state $\ket{0}$, and $\ket{\phi(t)}=-\ket{1}$ for an initial state $\ket{1}$. These operations may differ from ideal X and Z operations by a local phase factor, which can be compensated through single qubit phase gates or pulse optimization techniques. Besides, those local phase make the three qubits target becomes an ideal Y target. By solving the two equations $\tau=\frac{2\pi}{\sqrt{4\alpha^2+\delta_{t1}^2}}$ and $\delta_{t1}=\frac{\pi}{2\tau}$ , and selecting an appropriate gate time $\tau$, the values of parameters $\delta_{t1}$ and $\alpha$ can be determined. Since the Hamiltonian for target 2 is identical to that of target 1, the same parameters can be applied to target 2 for building an CYY or CXX gate. Besides, as the dynamics of the two qubits are independent, the operation for target 1 and 2 can be chosen arbitrary. By assigning X(Y) or Z operations, any gates with  X or Z target can be realized. \par
Similarly, the effective Hamiltonian of the five qubits model can be written as \begin{equation}\begin{aligned}
H&=\alpha X_{t1}+\alpha'Y_{t1}+\beta X_{t2} +\beta' Y_{t2}+\gamma X_{t3}+\gamma' Y_{t3}\\&+\delta_1 Z_{c1}X_{t1}+\delta_2 Z_{c1}X_{t2} +
\delta_3 Z_{c1}X_{t3}+\delta_4 Z_{c2}X_{t1}\\&+ \delta_5 Z_{c2}X_{t2} +\delta_6 Z_{c2}X_{t3}
\end{aligned}\end{equation}
 Under the four control states and applying the parameter condition $\alpha=-\delta_1=-\delta_4$, the Hamiltonian reduces to:
\begin{equation}\begin{aligned}
&H^{00}=-\alpha X+\alpha' Y\\
&H^{01}=\alpha X+\alpha' Y\\
&H^{10}=\alpha X+\alpha' Y\\
&H^{11}=3\alpha X+\alpha' Y
\end{aligned}\end{equation} for each target qubit(they can be calculated individually) the evolution of each qubit is governed by a time independent unitary operator $U=e^{-iHt}$,
\begin{equation}\begin{aligned}
&\ket{\phi(t)}=\\&
\text{cos}(\sqrt{\alpha^2+\alpha'^2}t)\ket{0}+\frac{-i\alpha+\alpha'}{\sqrt{\alpha^2+\alpha'^2}}\text{sin}(\sqrt{\alpha^2+\alpha'^2}t)\ket{1}\\&
\ket{\phi(t)}=\\&
\frac{-i\alpha-\alpha'}{\sqrt{\alpha^2+\alpha'^2}}\text{sin}(\sqrt{\alpha^2+\alpha'^2}t)\ket{0}+\text{cos}(\sqrt{\alpha^2+\alpha'^2}t)\ket{0}
\end{aligned}\end{equation} for initial state $\ket{0}$ and state $\ket{1}$ in the first three control conditions, \begin{equation}\begin{aligned}
&\ket{\phi(t)}=\\&
\text{cos}(\sqrt{9\alpha^2+\alpha'^2}t)\ket{0}+\frac{-i3\alpha+\alpha'}{\sqrt{9\alpha^2+\alpha'^2}}\text{sin}(\sqrt{9\alpha^2+\alpha'^2}t)\ket{1}\\&
\ket{\phi(t)}=\\&
\frac{-i3\alpha-\alpha'}{\sqrt{9\alpha^2+\alpha'^2}}\text{sin}(\sqrt{9\alpha^2+\alpha'^2}t)\ket{0}+\text{cos}(\sqrt{9\alpha^2+\alpha'^2}t)\ket{0}
\end{aligned}\label{r6}\end{equation} for initial state $\ket{0}$ and state $\ket{1}$ in the last control condition, the results demonstrate that under the first three conditions, the system oscillates at a Rabi frequency of $\sqrt{\alpha^2+\alpha'^2}$, while under the fourth condition, it oscillates at $\sqrt{9\alpha^2+\alpha'^2}$. This enables the implementation of an X operation for the control state $\ket{00}$, and the iCCXXX gate can be realized when an appropriate gate time $\tau$ is selected.

\section{Calculation for effective Hamiltonian by a way of Block Diagonalization}
\label{Appen_Bloch}
In this section, we present an alternative approach for
computing effective interactions based on block diagonal-
ization. 
To illustrate
the application of this method, we employ the three qubit
CXX model as a concrete example and we take three leval into account. The initial Hamiltonian:
\begin{equation}\begin{aligned}
H=&\frac{\omega_{t}}{2}b_{t1}^{\dagger}b_{t1} +\frac{\omega_{t}}{2}b_{t2}^{\dagger}b_{t2}+\frac{\omega_{c}}{2}b_{c}^{\dagger}b_{c}+\frac{\alpha_{t1}}{2}b_{t1}^{\dagger}b_{t1}(b_{t1}^{\dagger}b_{t1}-1)\\&+\frac{\alpha_{t2}}{2}b_{t2}^{\dagger}b_{t2}(b_{t2}^{\dagger}b_{t2}-1)+\frac{\alpha_{c}}{2}b_{c}^{\dagger}b_{c}(b_{c}^{\dagger}b_{c}-1)
+g_1(b_{c}^{\dagger}\\&b_{t1}+b_{c}b_{t1}^\dagger)+g_2(b_{c}^{\dagger}b_{t2}+b_{c}b_{t2}^\dagger)+\Omega_{t1}\text{cos}(\omega_{t}t+\phi_{t1})(\\&b_{t1}^{\dagger}+b_{t1})+\Omega_{t2}\text{cos}(\omega_{t}t+\phi_{t2})(b_{t2}^{\dagger}+b_{t2})+\Omega_{c}\text{cos}(\omega_{t}t\\&+\phi_{c})(b_{c}^{\dagger}+b_{c})
\end{aligned}\label{rr1}\end{equation}
and we take a RWA to this initial hamiltonian for ignoring those time dependent terms,
\begin{equation}\begin{aligned}
H_{RWA}=& \frac{\delta_{c}}{2}b_{c}^{\dagger}b_{c}+\frac{\alpha_{t1}}{2}b_{t1}^{\dagger}b_{t1}(b_{t1}^{\dagger}b_{t1}-1)+\frac{\alpha_{t2}}{2}b_{t2}^{\dagger}b_{t2}\\&(b_{t2}^{\dagger}b_{t2}-1)+\frac{\alpha_{c}}{2}b_{c}^{\dagger}b_{c}(b_{c}^{\dagger}b_{c}-1)
+g_1(b_{c}^{\dagger}b_{t1}\\&+b_{c}b_{t1}^\dagger)+g_2(b_{c}^{\dagger}b_{t2}+b_{c}b_{t2}^\dagger)+\frac{\Omega_{t1}}{2} (\text{e}^{-i\phi_{t1}}b_{t1}^{\dagger}\\&+\text{e}^{i\phi_{t1}}b_{t1})+\frac{\Omega_{t2}}{2}(\text{e}^{-i\phi_{t2}}b_{t2}^{\dagger}+\text{e}^{i\phi_{t2}}b_{t2})+\\&\frac{\Omega_{c}}{2} (\text{e}^{-i\phi_{c}}b_{c}^{\dagger}+\text{e}^{i\phi_{c}}b_{c})
\end{aligned}\end{equation}\par
The exact block diagonalization is normally calculated through the following steps. Firstly, we calculated the eigenvector matrix S of the rotating wave approximation (RWA) Hamiltonian. And then the block diagonalized
effective Hamiltonian can then be calculated using this
framework.
\begin{equation}H_{BD}=T^{\dagger}HT , T=SS_{BD}^{\dagger}(S_{BD}S^{\dagger}_{BD})^{-1/2}\end{equation}
with S represents the eigenvector matrix of the RWA Hamiltonian and $S_{BD}$ represents the block diagonalized matrix of matrix S.
Finally, the effective interactions can be calculated by trace the effective Hamiltonian and the interaction matrix.\par
And the Bloch-Brandow Perturbation following those steps. The RWA Hamiltonian can be written into $H=H_0+V$, them the effective Hamiltonian can be written as $H_{eff}=PH_0P+V_{eff}$, with P represents the projection operator to project the matrix into the computational subspace. And the $V_{eff}$ can be calculated through perturbation expansion.

\begin{equation}
\begin{aligned}
   & V_{eff}^{(1)}=PVP \\
   & V_{eff}^{(2)}=P[V(V)]P \\
   & ......
\end{aligned}
\end{equation} with the  superoperator (V)
\begin{equation}[(V)]_{Ii}=\frac{1}{\xi_i-\xi_I}\bra{I}V\ket{i}\label{rr5}\end{equation} with $\xi_i$,$\xi_I$ and $\ket{i}$,$\ket{I}$ represents eigenvalue and eigen states. So by following those steps, we can calculate the effective Hamiltonian in the two level computational subspace. Finally, by computing the trace over interactions formulated in Pauli form, we obtain all effective
interaction strengths.(we take the parameters condition that satisfy the CR effect)
\begin{widetext}
\begin{equation}
\begin{aligned}
XZI=& \frac{g_{\text{1}}}{4 \left(\alpha _c+\Delta _c\right){}^2} [\frac{2 \Omega _{t1} g_{\text{1}} \cos \left(\phi _{t1}\right) \left(\alpha _{t1}^3-2 \alpha _{t1}^2 \Delta _c+\alpha _{t1} \left(4 \alpha _c \Delta _c+2 \alpha _c^2+3 \Delta _c^2\right)-\Delta _c \left(\alpha _c+\Delta _c\right){}^2\right)}{\alpha _{t1} \left(\alpha _{t1}-\Delta _c\right){}^2}\\&+2 \left(\alpha _c+\Delta _c\right) \left(\frac{\Omega _{t2} g_{\text{c2}} \cos \left(\phi _{t2}\right) \left(-2 \alpha _{t2}-\alpha _c+\Delta _c\right)}{\alpha _{t2} \left(\alpha _{t2}-\Delta _c\right)}+2 \Omega _c \cos \left(\phi _c\right)\right)]
\end{aligned}
\label{rrr1}
\end{equation}

\begin{equation}
\begin{aligned}
ZZI=\frac{e^{-i \left(\phi _c+\phi _{t1}\right)} g_{\text{1}} \left(\Omega _{t1} \Omega _c \left(e^{2 i \phi _c}+e^{2 i \phi _{t1}}\right) \left(\alpha _{t1} \Delta _c+\left(\alpha _{t1}+\alpha _c\right){}^2-\Delta _c^2\right)-4 \alpha _{t1} \left(\alpha _{t1}+\alpha _c\right) e^{i \left(\phi _c+\phi _{t1}\right)} g_{\text{1}} \left(\alpha _c+\Delta _c\right)\right)}{4 \alpha _{t1} \left(\alpha _{t1}-\Delta _c\right) \left(\alpha _c+\Delta _c\right){}^2}
\end{aligned}
\end{equation}
\begin{equation}
\begin{aligned}
XZZ=\frac{\Omega _{t2} g_{\text{1}} g_{\text{2}} \cos \left(\phi _{t2}\right) \left(\alpha _{t2} \Delta _c+\left(\alpha _{t2}+\alpha _c\right){}^2-\Delta _c^2\right)}{2 \alpha _{t2} \left(\alpha _{t2}-\Delta _c\right) \left(\alpha _c+\Delta _c\right){}^2}
\end{aligned}
\label{rr2}\end{equation}
\begin{equation}
\begin{aligned}
......
\end{aligned}
\end{equation}

\end{widetext}
To illustrate the relative magnitudes of these interactions, we consider a representative three-qubit parameter set with \(\omega_c/2\pi=5.3\) GHz, \(\omega_{t1}/2\pi=\omega_{t2}/2\pi=\omega_d/2\pi=5.2\) GHz, \(\alpha_c/2\pi=\alpha_{t1}/2\pi=\alpha_{t2}/2\pi=-300\) MHz, \(g_1/2\pi=g_2/2\pi=8\) MHz, \(\Omega_c/2\pi=50\) MHz, and \(\Omega_{t1}/2\pi=\Omega_{t2}/2\pi=2\) MHz, with all drive phases set to zero. Eqs.~(\ref{rrr1})-(\ref{rr2}) predict \(|J_{XZI}|/2\pi=2.0001\) MHz, \(|J_{ZZI}|/2\pi=0.5067\) MHz, and \(|J_{XZZ}|/2\pi=4.267\) kHz, where \(J_P\) denotes the coefficient multiplying the Pauli operator \(P\). Thus, the additional two-body \(ZZI\) and three-body \(XZZ\) terms are approximately \(25.33\%\) and \(0.213\%\) of the desired control–target \(ZX\) interaction, respectively. Both terms are absent from the simplified QFR Hamiltonian in Eqs.~(\ref{222})-(\ref{333}), but are retained in the multilevel treatment presented here. The \(ZZI\) term produces conditional phases, whereas the \(XZZ\) term introduces a rotation conditioned on the joint state of two other qubits; both can therefore modify the intended gate evolution. Their effects on gate fidelity depend on the gate duration, pulse envelopes, and combined dynamics, and cannot be inferred directly from the strength ratios. This comparison illustrates how block diagonalization identifies additional interactions that should be assessed when evaluating the simplified effective description.
\section{Error Budget} \label{Appen_error}In this part we illustrate the details in error budget for our quantum system with the method in \cite{2026Echo} to show the detail relationship between the infidelity and control parameters. Although we have known the error budget from Eq.~(\ref{eq:example1}) and Eq.(\ref{eq:example2}) in the main text, we sometimes want to know the analytical form of the error budget for better understanding the relationship between our control pulse and the infidelity. And we take this regime into account. The errors in our system are generally divided into simple error which can be eliminated just by several single gate in the quantum circuit and is neglected here, leakage error , incoherent error , coherent error and the others. 

And the error caused by simply error can be estimated  by setting a matrix which is close to the real evolution unitary.~\cite{2019Operation} The fidelity of these two matrix can be written as \begin{equation}\begin{aligned}
F_{RU}=\frac{1}{n(n+1)}[Tr(RR\dagger)+Tr((R\dagger U)|^2]
\end{aligned}\label{rr3}\end{equation} and here we take a a matrix which is close to real unitary R and allow any rotation and quantum leap, \begin{equation}\begin{aligned}
U=&\text{e}^{i\theta_0}\ket{0}_{c}\bra{0}e^{-i(\phi_0/2)X_{t1}}+\text{e}^{i\theta_1}\ket{1}_{c}\bra{1}\text{e}^{-i(\phi_1/2)X_{t1}}+\\&
\text{e}^{i\theta_2}\ket{0}_{c}\bra{0}\text{e}^{-i(\phi_2/2)X_{t2}}+\text{e}^{i\theta_3}\ket{1}_{c}\bra{1}\text{e}^{-i(\phi_3/2)X_{t2}}
\end{aligned}\end{equation} and we take $U_0$ as the ideal unitary, so the simple error can be estimated as $(1-F(R,U_0))-(1-F(R,U))$, but until now ,we still don't know the parameters in U, and as U is the closed matrix to R ,we only need to make $R\dagger U$ to be maximum. The calculation for $R\dagger U$ is trivial, and after that one can found by taking the parameters below, the formula will be maxinum. \begin{equation}\begin{aligned}
&\phi_0=-\text{arg}\\&(\frac{R_{11}+R_{22}+R_{33}+R_{44}+R_{13}+R_{31}+R_{24}+R_{42}}{R_{11}+R_{22}+R_{33}+R_{44}-R_{13}-R_{31}-R_{24}-R_{42}})\\&\theta_0=\text{arg}[(R_{11}+R_{22}+R_{33}+R_{44})\text{cos}(\frac{\phi_0}{2})\\&+i(R_{13}+R_{31}+R_{24}+R_{42})\text{sin}(\frac{\phi_0}{2})]\end{aligned}\end{equation}
\begin{equation}\begin{aligned}
\\&\phi_1=-\text{arg}\\&(\frac{R_{55}+R_{66}+R_{77}+R_{88}+R_{57}+R_{75}+R_{68}+R_{86}}{R_{55}+R_{66}+R_{77}+R_{88}-R_{57}-R_{75}-R_{68}-R_{86}})\\&\theta_1=\text{arg}[(R_{55}+R_{66}+R_{77}+R_{88})\text{cos}(\frac{\phi_1}{2})\\&+i(R_{57}+R_{75}+R_{68}+R_{86})\text{sin}(\frac{\phi_1}{2})]\end{aligned}\end{equation}
\begin{equation}\begin{aligned}
\\&\phi_2=-\text{arg}\\&(\frac{R_{11}+R_{22}+R_{33}+R_{44}+R_{12}+R_{21}+R_{34}+R_{43}}{R_{11}+R_{22}+R_{33}+R_{44}-R_{12}-R_{21}-R_{34}-R_{43}})\\&\theta_2=\text{arg}[(R_{11}+R_{22}+R_{33}+R_{44})\text{cos}(\frac{\phi_2}{2})\\&+i(R_{12}+R_{21}+R_{34}+R_{43})\text{sin}(\frac{\phi_2}{2})]\end{aligned}\end{equation}
\begin{equation}\begin{aligned}
\\&\phi_3=-\text{arg}\\&(\frac{R_{55}+R_{66}+R_{77}+R_{88}+R_{56}+R_{65}+R_{78}+R_{87}}{R_{55}+R_{66}+R_{77}+R_{88}-R_{56}-R_{65}-R_{78}-R_{87}})\\&\theta_3=\text{arg}[(R_{55}+R_{66}+R_{77}+R_{88})\text{cos}(\frac{\phi_3}{2})\\&+i(R_{56}+R_{65}+R_{78}+R_{87})\text{sin}(\frac{\phi_3}{2})]
\end{aligned}\end{equation}
And now we can focus on the calculation for leakage error and coherent error, we then define a matrix R' which is close to U but only allow the rotation of target qubits.
\begin{equation}\begin{aligned}R'=& \ket{0}_{c}\bra{0}\text{e}^{-i(\phi^{'}_0/2)X_{t1}}+ \ket{1}_{c}\bra{1}\text{e}^{-i(\phi_1^{'}/2)X_{t1}}+\\&
 \ket{0}_{c}\bra{0}\text{e}^{-i(\phi_2^{'}/2)X_{t2}}+ \ket{1}_{c}\bra{1}\text{e}^{-i(\phi_3^{'}/2)X_{t2}}\end{aligned}\end{equation} and after to make $R\dagger R'$ to be max, we will find that $\phi^{'}_i=\phi_i$, then leakage error will be $(1-F(RR'))$ and coherent error will be $(1-F(R'U))$,

\begin{equation}
\begin{aligned}
1-F(U,R)=(1-F(R,R'))+(1-F(R',U))+\epsilon_{other}
\end{aligned}\label{rr4}
\end{equation} 
 with $\epsilon_{other}$ representing the other error.

\section{Optimization for verifying experiment feasibility and decomposition of multiqubit gate}
\label{Appen_opt}
\begin{figure}[t]
\centering
\includegraphics[width=8.7cm ]{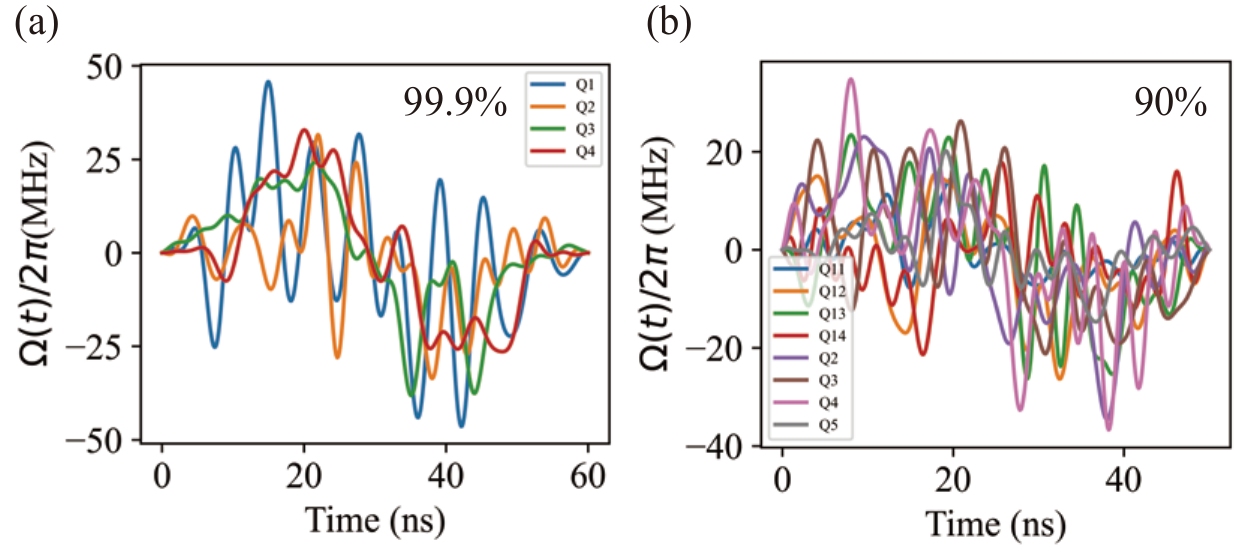}
\caption{Optimization for verifying experiment feasibility}
\label{opt2}
\end{figure}
The parameters for three qubits model are $\omega_{c}=5.5GHz$,$\omega_{t1}=5.2GHz$ and $\omega_{t2}=5GHz$,and two CR pulse at frequency $\omega_{t1}$ and $\omega_{t2}$ are applied on the control qubit with two pulse resonated with the target are applied on the two targets separately. The figure show the optimized shape of those pulses. The five qubits model take the similar regime with the five frequencies 5.5,5.3,5.2,5.1 and 5 GHz. The fidelity for three qubit model reaches 99.9\% and 90\% for five qubits model. Those results shows that the frequency crowded can be solved and show the experiment friendly of our regime. The robustness of our optimization results against the choice of optimization algorithm is further confirmed by performing additional optimizations using the GRAPE algorithm, which achieves comparable fidelities under the same conditions (see Table~\ref{com}).\par
To quantitatively characterize the cross-interaction between the two CR channels, we separately switch off one CR drive and compare the desired $ZX$ interaction on its corresponding target qubit with the unwanted off-resonant $ZX$ interaction induced on the other target qubit. Specifically, when CR2 (CR1) is switched off, we define the relative strength of the cross-interaction as
\begin{equation}
\eta_1=\left|\frac{\tilde{J}_{ZX}^{(1)}}{J_{ZX}^{(1)}}\right|,\qquad
\eta_2=\left|\frac{\tilde{J}_{ZX}^{(2)}}{J_{ZX}^{(2)}}\right|,
\end{equation}
where $J_{ZX}^{(i)}$ denotes the desired interaction and $\tilde{J}_{ZX}^{(i)}$ denotes the corresponding cross-interaction. For the parameters used in our simulations, both ratios are on the order of $10^{-1}$, indicating that the unwanted cross-interactions are approximately one order of magnitude weaker than the intended interactions. This provides a direct quantitative characterization of the relatively weak cross-interaction between the two CR channels.\par
\begin{figure}[t]
\centering
\includegraphics[width=8.5cm ]{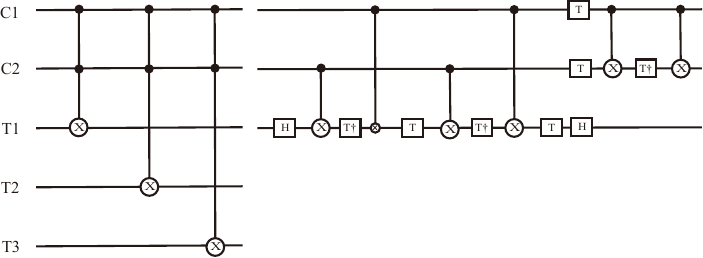}
\caption{Decomposition of CCXXX gate}
\label{opt5}
\end{figure}

And Fig.~\ref{opt5} shows the decomposition of our CCXXX gate to show the comparison between single step gate and decomposed one. The CCXXX can be decomposed into three Toffoli gates and the standard decomposition of Toffoli gate are shown on the right.

\begin{table}[t]
\caption{Comparison of gate infidelities optimized by COCOA and GRAPE.}
\label{com}
\centering
\centering
\renewcommand{\arraystretch}{1.8}
\begin{tabular}{c|cc}
\hline
Gate & COCOA & GRAPE \\
\hline
CYY   & $4.0\times10^{-3}$ & $6.0\times10^{-3}$ \\
CXX   & $4.0\times10^{-3}$ & $3.0\times10^{-3}$ \\
CZX   & $3.0\times10^{-3}$ & $5.0\times10^{-3}$ \\
CXXXX & $7.9\times10^{-2}$ & $9.2\times10^{-2}$ \\
\hline
\end{tabular}
\end{table}

\bibliography{reference}

\end{document}